\documentclass[fleqn,usenatbib]{mnras}

\usepackage{newtxtext,newtxmath}

\usepackage[T1]{fontenc}

\DeclareRobustCommand{\VAN}[3]{#2}
\let\VANthebibliography\thebibliography
\def\thebibliography{\DeclareRobustCommand{\VAN}[3]{##3}\VANthebibliography}

\usepackage{graphicx,graphics}	
\usepackage{amsmath}	
\usepackage{amssymb}	
\usepackage{bm}
\usepackage{xspace}

\newcommand{\wv}{\bm{w}}
\newcommand{\Wv}{\bm{W}}
\newcommand{\wvz}{\bm{w}(0)}
\newcommand{\Pv}{\bm{P}}
\newcommand{\Rv}{\bm{R}}

\newcommand{\diff}{\mathrm{d}}
\newcommand{\dt}{\Delta t}
\newcommand{\dEoE}{|\diff E/E|}
\newcommand{\pt}{p_{\mathrm{t}}}
\newcommand{\tbin}{T_{\mathrm{bin}}}
\newcommand{\tbinin}{T_{\mathrm{bin,in}}}
\newcommand{\rh}{R_{\mathrm{h}}}
\newcommand{\tr}{T_{\mathrm{r}}}
\newcommand{\trh}{T_{\mathrm{rh}}}
\newcommand{\tcr}{T_{\mathrm{cr}}}
\newcommand{\ma}{m_{\mathrm{1}}}
\newcommand{\mb}{m_{\mathrm{2}}}
\newcommand{\mk}{m_{k}}
\newcommand{\mi}{m_{i}}
\newcommand{\mj}{m_{j}}
\newcommand{\ab}{a_{\mathrm{h/s}}}
\newcommand{\mave}{\langle m \rangle}

\newcommand{\Nb}{N_{\mathrm{b}}}
\newcommand{\Na}{N_{\mathrm{act}}}
\newcommand{\Nbp}{N_{\mathrm{b,p}}}
\newcommand{\Hsd}{H_\mathrm{sd}}
\newcommand{\Hb}{H_{\mathrm{b}}}
\newcommand{\vi}{\bm{v}_i}

\newcommand{\ms}{m_{\mathrm s}}
\newcommand{\mpr}{m_{\mathrm p}}

\newcommand{\Cs}{C_{\mathrm{s}}}
\newcommand{\nbfac}{C_{\mathrm{r}}}
\newcommand{\nbvfac}{C_{\mathrm{v}}}

\newcommand{\HL}{H_{\mathrm{L}}}
\newcommand{\HS}{H_{\mathrm{S}}}
\newcommand{\DL}{\mathcal{L}}
\newcommand{\DS}{\mathcal{S}}
\newcommand{\pppt}{$\mathrm{P^3T}$}
\newcommand{\pkick}{$\mathrm{P^3T}$-kick}
\newcommand{\pdrift}{$\mathrm{P^3T}$-drift}
\newcommand{\pppm}{$\mathrm{P^3M}$}

\newcommand{\rin}{r_{\mathrm{in}}}
\newcommand{\rout}{r_{\mathrm{out}}}
\newcommand{\rini}{r_{\mathrm{in},i}}
\newcommand{\routi}{r_{\mathrm{out},i}}
\newcommand{\rinj}{r_{\mathrm{in},j}}
\newcommand{\routj}{r_{\mathrm{out},j}}
\newcommand{\rinr}{r_{\mathrm{in,ref}}}
\newcommand{\routr}{r_{\mathrm{out,ref}}}
\newcommand{\rinij}{r_{\mathrm{in},ij}}
\newcommand{\routij}{r_{\mathrm{out},ij}}
\newcommand{\rcmk}{r_{\mathrm{cm},k}}
\newcommand{\rink}{r_{\mathrm{in},k}}

\newcommand{\rnbi}{r_{\mathrm{nb},i}}
\newcommand{\rpij}{r_{\mathrm{p},ij}}
\newcommand{\rij}{r_{ij}}
\newcommand{\WU}{f}
\newcommand{\WF}{K}
\newcommand{\dts}{\Delta t_{\mathrm{L}}}
\newcommand{\dthi}{\Delta t_{\mathrm{H},i}}
\newcommand{\dthm}{\Delta t_{\mathrm{H,max}}}
\newcommand{\fr}{f_{\mathrm{p}}}
\newcommand{\rgij}{r_{\mathrm{g},ij}}

\newcommand{\petar}{\textsc{petar}}
\newcommand{\petarurl}{https://github.com/lwang-astro/PeTar}
\newcommand{\nbpp}{\textsc{nbody6++gpu}}
\newcommand{\fdps}{\textsc{fdps}\xspace}
\newcommand{\sdar}{\textsc{sdar}\xspace}

\newcommand{\tcomp}{T_{\mathrm{W}}}

\newcommand{\Ncore}{N_{\mathrm{core}}}

\title[PeTar]{PeTar: a high-performance $N$-body code for modeling massive collisional stellar systems}

\author[Long Wang et al.]{
Long Wang,$^{1,2}$\thanks{E-mail:long.wang@astron.s.u-tokyo.ac.jp}
Masaki Iwasawa,$^{2,3}$
Keigo Nitadori$^{2}$
and Junichiro Makino$^{2,4}$
\\
  $^{1}$Department of Astronomy, School of Science, The University of Tokyo, 7-3-1 Hongo, Bunkyo-ku, Tokyo, 113-0033, Japan \\
  $^{2}$RIKEN Center for Computational Science, 7-1-26 Minatojima-minami-machi, Chuo-ku, Kobe, Hyogo 650-0047, Japan\\
  $^{3}$National Institute of Technology, Matsue College, 14-4, Nishi-ikuma-cho, Matsue, Shimane 690-8518, Japan \\
  $^{4}$Graduate School of Science, Kobe University, 1-1 Rokkodai-cho, Nada-ku, Kobe, Hyogo 657-8501, Japan \\
}

\date{Accepted XXX. Received YYY; in original form ZZZ}

\pubyear{2020}

\begin{document}
\label{firstpage}
\pagerange{\pageref{firstpage}--\pageref{lastpage}}
\maketitle

\begin{abstract}
  The numerical simulations of massive collisional stellar systems, such as globular clusters (GCs), are very time-consuming.
  Until now, only a few realistic million-body simulations of GCs with a small fraction of binaries ($5\%$) have been performed by using the \nbpp~code.
  Such models took half a year computational time on a GPU based supercomputer.
  In this work, we develop a new $N$-body code, \petar, by combining the methods of Barnes-Hut tree, Hermite integrator and slow-down algorithmic regularization (SDAR).
  The code can accurately handle an arbitrary fraction of multiple systems (e.g. binaries, triples) while keeping a high performance by using the hybrid parallelization methods with MPI, OpenMP, SIMD instructions and GPU.
  A few benchmarks indicate that \petar~and \nbpp~have a very good agreement on the long-term evolution of the global structure, binary orbits and escapers.
  On a highly configured GPU desktop computer, the performance of a million-body simulation with all stars in binaries by using \petar~is $11$ times faster than that of \nbpp.
  Moreover, on the Cray XC50 supercomputer, \petar~well scales when number of cores increase. The ten million-body problem, which covers the region of ultra compact dwarfs and nuclear star clusters, becomes possible to be solved.
\end{abstract}




\section{Introduction}

The realistic models of collisional stellar systems where stars and binaries can have frequent close interactions, such as globular clusters (GCs), have been a long-term challenge due to the time consuming calculations.
By developing a hybrid-parallel direct $N$-body code \nbpp~\citep{Wang2015}, the million-body models of GCs have become possible \citep[DRAGON models;][]{Wang2016}.
This success is based on the breakthrough of both hardware and software developments.
Especially, the Gravity Pipe \citep[GRAPE;][]{Makino2003} and the Graphic Processing Unit (GPU) provide great platforms for performing highly-efficient parallelized simulation codes \citep{Gaburov2009}.

However, the DRAGON models have relatively lower densities compared to the typical GCs in the Galaxy.
It is difficult to simulate high density systems because the half-mass relaxation timescale ($T_\mathrm{rh}$) is much shorter, while the calculation cost scales in a way of $O(N^3)/T_\mathrm{rh}$.
Even so, the DRAGON models took about half a year computing to reach one $T_\mathrm{rh}$.
Besides, the binary fraction is also small ($5\%$) compared to that in the observed GCs.
This is because the orbital integration of binaries is not parallelized in \nbpp, thus the performance cannot well scale with multiple CPU cores when the binary fraction is large.
Therefore, the dense models with many binaries are still challenging.
It is also not practicable to perform a large number of models to cover the parameter spaces of different initial conditions.
All these limit the applications of the direct $N$-body methods for studying GCs.

On the other hand, the ultra compact dwarfs (UCDs) attract researchers' attention to understand their formation and evolution, because they belong to a family crossing the region of GCs and dwarf galaxies \citep[e.g][]{Drinkwater2000,Hilker1999,Phillipps2001}.
The discovery of supermassive black hole (SMBH) in a UCD \citep{Seth2014} brings a new question that how SMBHs and UCDs co-evolve.
In the view of stellar dynamics, UCDs with $N\sim 10^7-10^8$ are in the transiting region between the collisional and collisionless systems.
Thus, a proper model of UCDs also requires a correct treatment of collisional dynamics.

The galactic nuclear star clusters (NSCs) are very massive and dense.
The collisional effect is also important.
They can be a few magnitude more massive than GCs but the system size is similar.
The formation scenarios of NSCs and the co-evolution among NSCs, SMBHs and host galaxies are still not fully understood.
The star-by-star simulations including SMBH are useful to study such kind of problems.
The currently most dense model of NSCs with a proper treatment of collisional effect was done by using the \nbpp~code \citep{Panamarev2019}, but the number of stars is still far from the realistic case.



To solve the challenge of realistically simulating these dense and massive systems, a new approach of numerical tool that can overcome the bottlenecks (performance and small fraction of binaries) is necessary.
The major difficulty is to solve the multiple timescale issue.
In this work, we describe how our new code, \petar, can overcome this challenge.

In Section~\ref{sec:timescale}, we introduce the multiple timescale issue.
In Section~\ref{sec:nb6} we summarise the algorithms used in the direct $N$-body methods, especially those in the \textsc{nbody6(++gpu)} code.
The approach of particle-tree with individual time steps is shortly described in Section~\ref{sec:ptits}.
Then we introduce the idea of Hamiltonian splitting in Section~\ref{sec:hsplit}.
The detailed description of \petar~is in Section~\ref{sec:petar}.
After that, we provide the benchmarks of \petar~in Section~\ref{sec:test}.
Finally, we make conclusions and discuss the future work in Section~\ref{sec:conclusion}.

\section{Multiple timescale issue}
\label{sec:timescale}

Long-term dynamical evolution of star clusters involves multiple physical processes with very different timescales.
Three important ones are:
\begin{itemize}
\item $\tbin$: period of binaries (order of day at the minimum).
\item $\tcr$: time for a star crossing the cluster (order of Myr for GCs).
\item $\tr$: two-body relaxation time of the system (order of Gyr for GCs).
\end{itemize}
The $N$-body method needs to be possible to handle these three timescale regions.

\subsection{Binary period}

Binaries contain very large mechanical energy compared to that of the host system.
They play an important role in controlling the global evolution via few-body interactions.
Since the gravitationally bound system has a negative heat capacity, the temperature of the cluster centre becomes hotter as energy is transferred outwards.
Such process is unstable and finally the core collapse happens and the central density increases significantly \citep[e.g.][]{Binney1987,Spitzer1987}.
Binaries are considered as the major heating source that can prevent an infinite core collapse.
After a close interaction between a tight binary and its neighbours in the centre, energy is added the system to support the core while the binary becomes more compact and finally merges or escapes from the system.
Thus, such process influences both the global evolution and the statistical properties of binaries.
This feature makes the star cluster an efficient environment to form exotic objects such as blue stragglers, X-ray binaries and gravitational wave progenitors.
Therefore, to obtain realistic models of star clusters, few-body interactions and dynamics of binaries must be correctly treated.

However, binaries also bring the major challenge for the $N$-body simulations.
$\tbin$ can cover a very wide region in star clusters.
The minimum of $\tbin$ is determined by the stellar structure of binary components.
Typically it is a few days.
There is no upper limit of $\tbin$, but in a cluster environment wide binaries are easily disrupted by perturbations.
Due to the Heggie-Hill law \citep{Heggie1975,Hills1975}, the boundary between the wide and the tight can be estimated as
\begin{equation}
  \ab \approx \frac{G \ma \mb}{\mave \sigma^2},
  \label{eq:ab}
\end{equation}
where $G$ is the gravitational constant; $\ma$ and $\mb$ are the masses of the two components, respectively; $\mave$ is the locally averaged stellar mass and $\sigma$ is the local velocity dispersion.
After close encounters, wide binaries with the semi-major axis $a>\ab$ become wider and while tight binaries become tighter.
Thus binaries with $a<\ab$ can stay in clusters for a long time before merger or escaping.

We can roughly estimate $\tbin$ at the boundary of $\ab$.
In Eq.~\ref{eq:ab}, $\sigma^2$ can be replaced by $G M/r$ for a system in virial equilibrium, where $M$ is the total mass and $r$ is the size of the system.
Thus, $\ab \approx r/N$ if all stars have a similar mass.
In an open cluster, $r\sim 1$~pc and $N\sim 10^3$, binaries close to the boundary have $\tbin$ in the order of $10^3$ years.
In a dense GC, the boundary $\tbin$ is in the order of years.
Thus, $\tbin$ has a range of $3-6$ orders of magnitudes.

To properly follow the orbital motion, the time step size of integration should be much less than $\tbin$.
If the orbit has a fast change at the peri-centre due to a high eccentricity ($e$), the time step should be very small to catch the peri-centre motion.
This is also the case for a hyperbolic encounter.
Thus, binaries and close encounters are the most time-consuming part in the $N$-body simulation of star clusters.

\subsection{Crossing time}

$\tcr$ represents the timescale of the orbital motion of a single star in the system.
Thus, the time step of integration for a star should be much shorter than its $\tcr$ to obtain a sufficient accuracy.
In a star cluster, $\tcr$ can be estimated by
\begin{equation}
  \tcr \approx \sqrt{\frac{1}{G\rho}},
\end{equation}
where $\rho$ is the local mean density.
After core collapse, the density contrast between the centre and the half-mass radius of star clusters can be an order of $10^4$ \citep[e.g.][]{Binney1987,Wang2016}.
Thus, $\tcr$ varies about $100$ times from the centre to the halo.
Even without binaries, the individual-time-step method is necessary to efficiently handle such a large range.

\subsection{Relaxation time}

Two-body relaxation is one important physical process that determines the long-term behaviour of the $N$-body system.
The phenomenons of core collapse, mass segregation and escaping of stars all depend on it.

The relation between $\tr$ and $\tcr$ can be described as \citep[e.g.][]{Binney1987}
\begin{equation}
  \tr \approx  \frac{0.1 N}{\ln \Lambda} \tcr ,
  \label{eq:tr}
\end{equation}
where $\ln \Lambda$ is Comlumb logarithm.
The factor $N$ in Eq.~\ref{eq:tr} indicates that for a global cluster with million stars, the ratio between $\tr$ and $\tcr$ is very large.
For a single-mass system in virial equilibrium, the averaged $\tr$ measured at the half-mass radius of the system ($\rh$) can be estimated by \citep{Spitzer1987}
\begin{equation}
  \trh \approx 0.138 \frac{N^{1/2} \rh^{3/2}}{\mave^{1/2} G^{1/2} \ln \Lambda},
  \label{eq:trh}
\end{equation}
Typically, GCs in the Galaxy have $\trh$ in an order of Gyr and already passed a few $\trh$.

To study the long-term evolution of star clusters, the numerical simulations need to cover at least one $\trh$.
However, the time resolution of the integration should be less than $\tbin$ and $\tcr$.
As the maximum of $\trh/\tbin\sim 10^{11}$, the classical integrators using individual-time-step methods (e.g. fourth-order Hermite with block time steps) is not practically possible to handle such expensive calculations.
Therefore, the sophisticated $N$-body codes (e.g. \textsc{nbody6(++gpu)}) apply special algorithms to reduce the computational operations.

\section{Direct N-body method}
\label{sec:nb6}

We provide a short review of the algorithms used in the direct $N$-body code, especially \textsc{nbody6(++gpu)} \citep{Aarseth2003,Nitadori2012,Wang2015}, that are designed to deal with the multiple timescale issue.
A part of the algorithms are also implemented in the \petar~code.

\subsection{Individual time steps}

The performance of force interaction calculation in direct $N$-body simulations are usually considered as $O(N^2)$ due to the pair interaction between all particles (stars).
However, this is only the case when the interaction between all particles are needed.
When the multiple timescale issue exists, sophisticated $N$-body codes for simulation star clusters like \textsc{nbody6(++gpu)} use individual time steps for each particle.
Thus particles with different $\tcr$ can use suitable integration steps to avoid expensive $O(N^2)$ calculations every step.
In \textsc{nbody6(++gpu)}, the fourth-order Hermite integrator with the block-time-step method is used.
The block-time-step method normalizes the step size to be an integer power of $0.5$, so that the implementation of multiple-core parallelization becomes possible.
The performance of interaction calculation per step is $O(N\langle \Na \rangle)$, where $\langle \Na \rangle$ is the averaged active particle number that need the update of forces at one step.
\cite{Makino1988} found that the total number of pair interactions depends on $O(N^{7/3})/\tcr$ if the system has a power-law density distribution with power-index $\alpha<24/11$.
For $\alpha>24/11$, the scaling relation depends on $\alpha$.
Using Eq.~\ref{eq:tr}, the scaling becomes $O(N^{10/3}/\ln \Lambda)/\tr$ for $\alpha<24/11$.
Thus, as $N$ increases, the computational cost grows rapidly.

\subsection{AC neighbour scheme}

To reduce the computation cost when $N$ is large, \cite{Ahmad1973} introduced the (AC) neighbour scheme, where the force on a particle is split into the short-distance (neighbour) part and long-distance (regular) part.
As the long-distance force changes smoothly, it can be updated with a larger time step (regular step) compared to that of the neighbour force.
Between two regular steps, the long-distance force is estimated by a second-order prediction.
Usually, the number of neighbour particles is a small fraction of total $N$ (the order of $10$--$10^2$), thus the total number of pair interactions is significantly reduced.
Especially, when small neighbour steps are required to handle close encounters and binaries, only $10$--$10^2$ force evaluations are needed per step while the regular step can be much larger.
Thus, the frequent $O(N)$ calculation is avoided.
The speed gained by the AC scheme is roughly proportional to $N^{1/4}$ without binaries \citep{Makino1988,Makino1992}.

If no short-period binaries exist, individual time steps combined with the AC neighbour scheme is an efficient method for simulating the long-term evolution of star clusters.
However, it is still not sufficient to handle the very large timescale gap caused by the short-period binaries.

\subsection{Binary integrator}

Since short-period binaries are very compact, the perturbation from neighbour particles is usually very weak unless a close encounter happens.
Based on this feature, \textsc{nbody6(++gpu)} codes do not evolve such binaries and treat them as single (centre-of-the-mass) particles, until the perturbation becomes strong enough.
These frozen binaries are named as ``isolated binaries'' in the codes.
Thus, only the internal motion of strongly perturbed binaries are actually integrated.
Besides, the time steps to integrate the internal motion are much smaller than the neighbour steps.
To avoid large number of pair interactions, only the force from nearby perturbers is included.
The typical number of such perturbers for a binary is less than $5$.
Therefore, the computational cost is reduced significantly.

The major perturber selection criterion is based on the strength of tidal force \citep[Equation 8.58;][]{Aarseth2003}:
\begin{equation}
  r_{\mathrm p} < \left[\frac{2m_{\mathrm p}}{M\gamma_{\mathrm{min}}}\right]^{1/3}R,
  \label{eq:tdp}
\end{equation}
where $r_{\mathrm p}$ is the distance between the perturber and the closest component in the binary;
$m_{\mathrm p}$ is the mass of the pertuber and $M$ is the mass of the binary;
$R$ is the apo-centre distance;
and $\gamma_{\mathrm{min}}$ is a free coefficient.

However, this criterion may ignore the impact from the whole system because it checks only the individual neighbours, but not the cumulative effect from the group of particles with the similar distances and directions.
For example, the central region of a GC can contain $10^5 M_\odot$.
If we consider its centre-of-mass $m_{\mathrm p} \approx 10^5 M_\odot$, the corresponding criterion $r_{\mathrm{p,G}}$ is $300$ times larger than the case ($r_{\mathrm{p,s}}$) of a normal star with $1 M_\odot$.
But by using this criterion, most stars outside $r_{\mathrm{p,s}}$ are excluded.
For the relative wide binaries in the outside region of a GC, this cumulative effect may be significant and cannot be ignored.
This is one potential problem when only close perturbers are selected.

On the other hand, to handle the highly eccentric binaries and close encounters, the \cite{Kustaanheimo1965} (KS) regularization and ``Algorithmic Regularization'' \citep[AR;][]{Mikkola1999} are used.
The regularization method avoids the singularity of Newtonian force when two particles get very close, thus no small time steps are required while the accuracy of integration is high enough.

%
%

\subsection{Slow-down algorithm}
\label{sec:slowdown}

\cite{Mikkola1996} developed the slow-down (SD) method that can significant reduce the number of integration steps for weakly perturbed binaries.
The key idea is to modify the Hamiltonian of a system with a binary as
\begin{equation}
  \Hsd = \frac{1}{\kappa} \Hb + (H - \Hb),
  \label{eq:sd}
\end{equation}
where $\Hsd$ is the new Hamiltonian and $\Hb$ is the Hamiltonian of the binary components.
The $\kappa$ is a scaling factor that slows down the motion of the binary that the effective period becomes $\tbin/\kappa$.
Thus the number of integration steps for this binary is also reduced by $\kappa$ times.
It is shown in \cite{Mikkola1996} that the secular motion of binary can be correctly reproduced while the orbital phase information is lost.

The \textsc{nbody6(++gpu)} codes apply the SD method together with the KS regularization.
However, since most weakly perturbed binaries are treated as isolated binaries, and a strict limit of $\kappa$ ($\le 10$) is used, the performance improvement is not significant.

\subsection{Parallelization}

With the algorithms described above, the total computational cost is still significant (roughly $O(N^3)/\trh$) for the direct $N$-body method.
Thus, the multiple-core parallelization is necessary to reduce the wall clock time of the computation.
\cite{Spurzem1999} and \cite{Hemsendorf2003} implemented the MPI parallelization for calculating the long-distant and neighbour force.
Then \cite{Nitadori2012} implemented the hybrid parallelization methods (\textsc{nbody6-gpu}) including the OpenMP, GPU (CUDA) and SIMD instructions (SSE, AVX) for long-distant and neighbour integration.
\cite{Wang2015} combined these two and optimize the code (\nbpp) in order to perform large $N$ simulations on supercomputers.

\subsection{Bottleneck}
\label{sec:bottleneck}

Although \nbpp~can handle the million-body simulations of GCs \citep{Wang2015}, there are several bottlenecks of the code that limit the future improvement.
Firstly, the large memory space is required to save the data.
For each particle, a neighbour list with few hundreds $4$-bytes integers needs to be saved.
Moreover, each MPI process keeps the complete copy of particle data.
Thus the maximum $N$ is limited by the maximum memory size per MPI process.
For example, if the maximum neighbour number is $500$, one million particles require $2$ GB memory space to save the neighbour lists.
Including the particle data and many others, the actual memory cost is significant.
Therefore, the code cannot be used to simulate systems with a very large $N$ even if the computing resource has a large number of computing nodes.

Secondly, the parallelization of integrating internal motions of binaries (KS regularization) is difficult.
One may think that this should not be since each binary can be evolved almost independently.
In reality, several parts that required a large number of operations are not possible to be parallized.
Especially, each time when the integrator for a binary needs to be switched between the KS regularization and the Hermite method, the initialization of the force requires a $O(N)$ pair interaction.
In addition, each particle has to update its neighbour list with the total memory access of $O(N\langle \Nb \rangle)$.
If switching is frequent, such cost is very large.
Unfortunately, there are always wide binaries that are close to the switching conditions in a star clusters if the primordial binary fraction is significant.

Moreover, the code is initially not designed for parallelization.
During the KS integration, many shared global variables (mostly for binary stellar evolution) are modified and conditional interruptions are frequently used.
Thus, the shared memory parallelization method like OpenMP is difficult to implement and does not scale with number of threads.
The thread safety is also not guaranteed.
On the other hand, the floating-point operation per memory access is low due to a few number of perturbers, thus the distributed memory parallel method (MPI parallelization) also does not scale \citep{Wang2015}.

Thirdly, the code mixes the $N$-body integration and the stellar evolution in a complex way.
It is not easy to separate the two parts.
Therefore, it is not flexible to replace the implementations of stellar-evolution models.
It is also challenging to maintain the code and include new features.


\section{Particle-tree with individual time steps}
\label{sec:ptits}

The force contributed by the distant particles are much weaker than that of neighbours.
Thus in the individual time step method, even the time-resolution is low for the weakly interacted particles, the accuracy is sufficient.
We can consider it as an approximation on time.
There are another type of $N$-body algorithms using an approximation on space, such as the Barnes-Hut tree \citep[PT;]{Barnes1986}, the particle mesh \citep[PM;]{Hockney1988} and the fast multiple method \citep[FMM;][]{Greengard1987}.
Compared to the cost of $O(N^2)$ in the direct force calculation of all particles, such methods only requires $O(N\log N)$ (PT and PM) or $O(N)$ (FMM).
However, these methods need a shared time step.
For example, the second-order leapfrog integrator is used in the PT method.
Therefore, it is difficult to use it for star clusters because of the multiple timescale issue.

\cite{McMillan1993} started the first effort to overcome this bottleneck by introducing a high-order predictor-corrector integrator with individual time steps in the PT method.
They also implemented the KS regularization to deal with binaries.
\cite{Fukushige2016} implemented the parellelization support on the GRAPE-9 system for this method and showed that it is much faster than the Hermite integrator for million-body simulations.

The PT method approximates the long-distant force.
However, the weak encounters from distant particles are important in the relaxation process.
Thus, it is necessary to ensure that the PT method can correctly reproduce the relaxation.
\cite{Hernquist1987} found that the accuracy of the relaxation process in the PT method depends on the opening angle ($\theta$).
When $\theta<1.0$, the measured $\trh$ of the PT method is consistent with that of the direct $N$-body method.

\section{Hybrid methods with Hamiltonian splitting}
\label{sec:hsplit}

Recently, hybrid numerical simulation methods become popular to solve the multiple timescale issues.
The key idea is based on the Hamiltonian splitting.
If the Hamiltonian the system can be decomposed to two parts as
\begin{equation}
  H = \HL + \HS
  \label{eq:split}
\end{equation}
The equation of motion can be described as
\begin{equation}
  \frac{\diff \wv}{\diff t} = \{ \wv, H\} = \{ \wv, \HL \} + \{ \wv, \HS \}.
\end{equation}
where $\{\}$ is Poisson bracket.
We define the differential operator $\DL \equiv \{\ , \HL\}$ and $\DS \equiv \{\ , \HS\}$.
Using matrix exponential, the symplectic mapping from $t$ to $t+\dt$ can be written as
\begin{equation}
  \wv(t+\dt) = e^{\dt(\DL+\DS)}\wv(t).
\end{equation}
If the two parts have analytic solutions, the symplectic integrator can be constructed.
The second-order symplectic integrator is given by
\begin{equation}
  \wv(t+\dt) = e^{\dt\DL/2}e^{\dt\DS}e^{\dt\DL/2}\wv(t) + O(\dt^3).
  \label{eq:lfmap}
\end{equation}
If $\HL$ and $\HS$ are the potential and kinetic energy, respectively,
Eq.~\ref{eq:lfmap} represents the leapfrog method with the order of kick-drift-kick.

The potential and kinetic energy are not the only combination.
\cite{Wisdom1991} firstly introduced the MVS method to evolve the planetary system, where $\HL$ and $\HS$ represent the Kepler motion and interactions between planets, respectively.
Then, several combinations of $\HL$ and $\HS$ are introduced to simulate different type of systems \citep[e.g.][]{Hockney1988,Xu1995,Chambers1999,Fujii2007,Oshino2011}.

\subsection{Particle-particle particle tree method}

For particle based $N$-body systems, one possible way of Hamiltonian splitting is using $\HL$ and $\HS$ to represent the long-range and short-range interactions, respectively.
Since $\HL$ dominates the computation while the contribute to the pair interaction is less than $\HS$, the approximated methods can be used with a large fixed time step, which provides a sufficient accuracy and a small computational cost.
On the other hand, more accurate methods with smaller and individual time steps can be applied for $\HS$.
These hybrid methods are used for several combinations, such as PM + PP  \citep[\pppm;][]{Hockney1988}, PM + PT \citep{Xu1995} and PT + PP \citep[\pppt;][]{Oshino2011}, where PP represents the direct $N$-body (particle-particle) method.


\begin{figure}
  \centering
  \includegraphics[width=0.8\columnwidth]{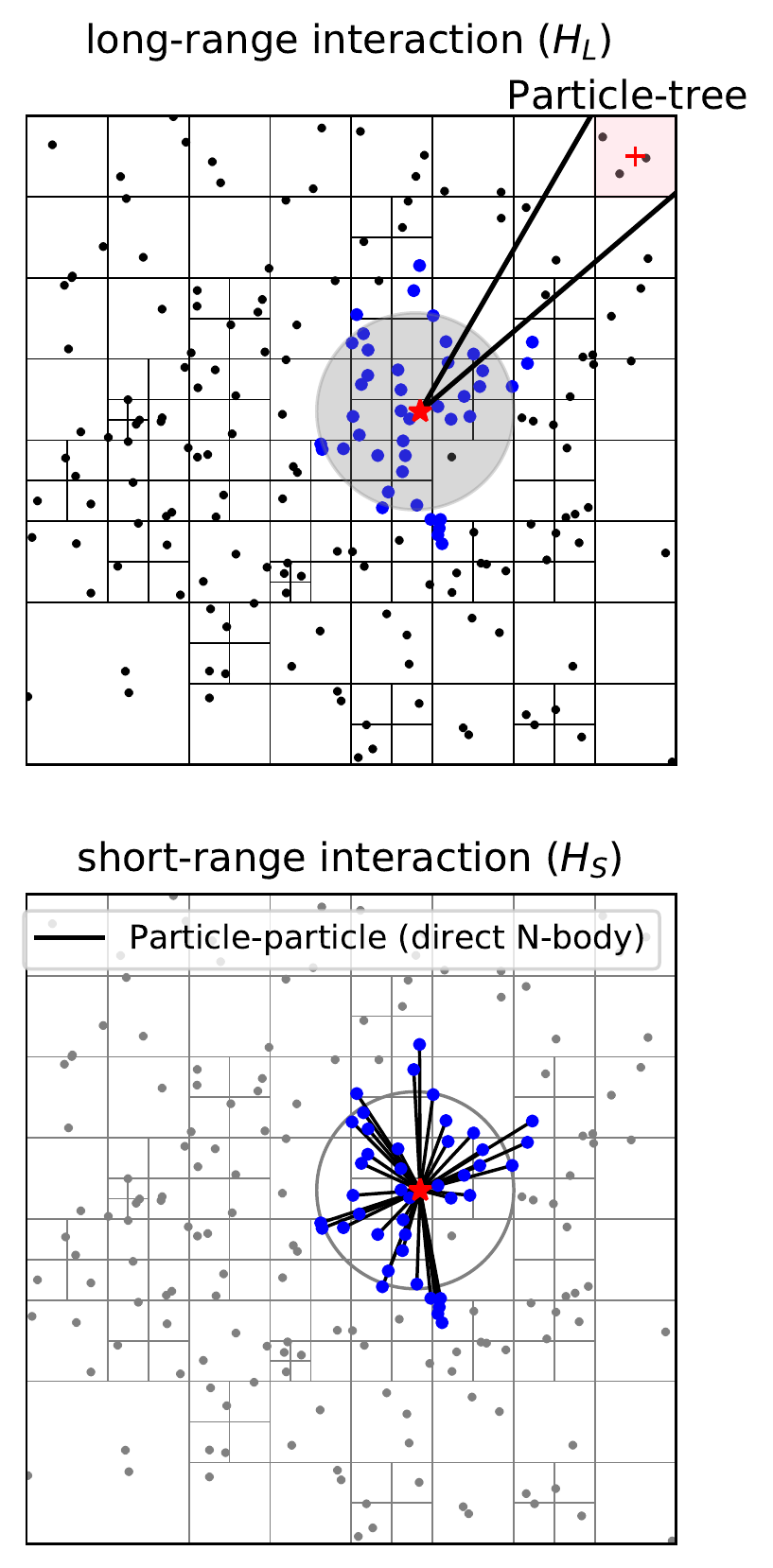}
  \caption{An illustration showing how the \pppt~method deal with the long-range and short-range interactions for a two-dimensional particle system.
    The upper panel shows the structure of the Barnes-Hut tree.
    The neighbour particles inside a distance criterion are collected as individual clusters. 
    An example for one (red) particle is shown as blue points.
    Two lines with an opening angle, $\theta=0.35$, starting from this particle are also shown.
    If a tree cell (like the pink box) is inside $\theta$, its super-particle (the centre-of-the-mass with multipole expansions) is used to obtain the long-range acceleration.
    The short-range acceleration for this particle (shown in the bottom panel) is calculated by the high-accuracy particle-particle (PP) method.
  }
  \label{fig:pppt}
\end{figure}

The \pppt~method introduced by \cite{Oshino2011} (Fig.~\ref{fig:pppt}) is specially designed to simulate the collisional systems which have the multiple timescale issue (without binaries).
The Hamiltonian splitting of this method is via a changeover function $W(\rij)$:
\begin{equation}
  \begin{aligned}
  \HS =& \sum_{i=1}^N\frac{p_i^2}{2 m_i} - \sum_{i<j}^N \frac{G m_i m_j}{\rij} W(\rij) \\
  \HL =& \sum_{=1}^N  \frac{G m_i m_j}{\rij} [1-W(\rij)] .\\
  \end{aligned}
  \label{eq:cf}
\end{equation}
The purpose of $W(\rij)$ is to result in a smooth transition when two particles pass the boundary of long-range and short-range interactions.


\cite{Iwasawa2015} developed a GPU-parallelized \pppt~code and compared its performance with the Hermite method.
They showed that the new scheme can be $10$ times faster.
This high performance encourages us to advance in this direction by combining the binary solver into the \pppt~method in order to properly handle the short-time interval close interactions.

\section{Hybrid N-body code: \petar}
\label{sec:petar}

\begin{figure*}
  \centering
  \includegraphics[width=1.0\textwidth]{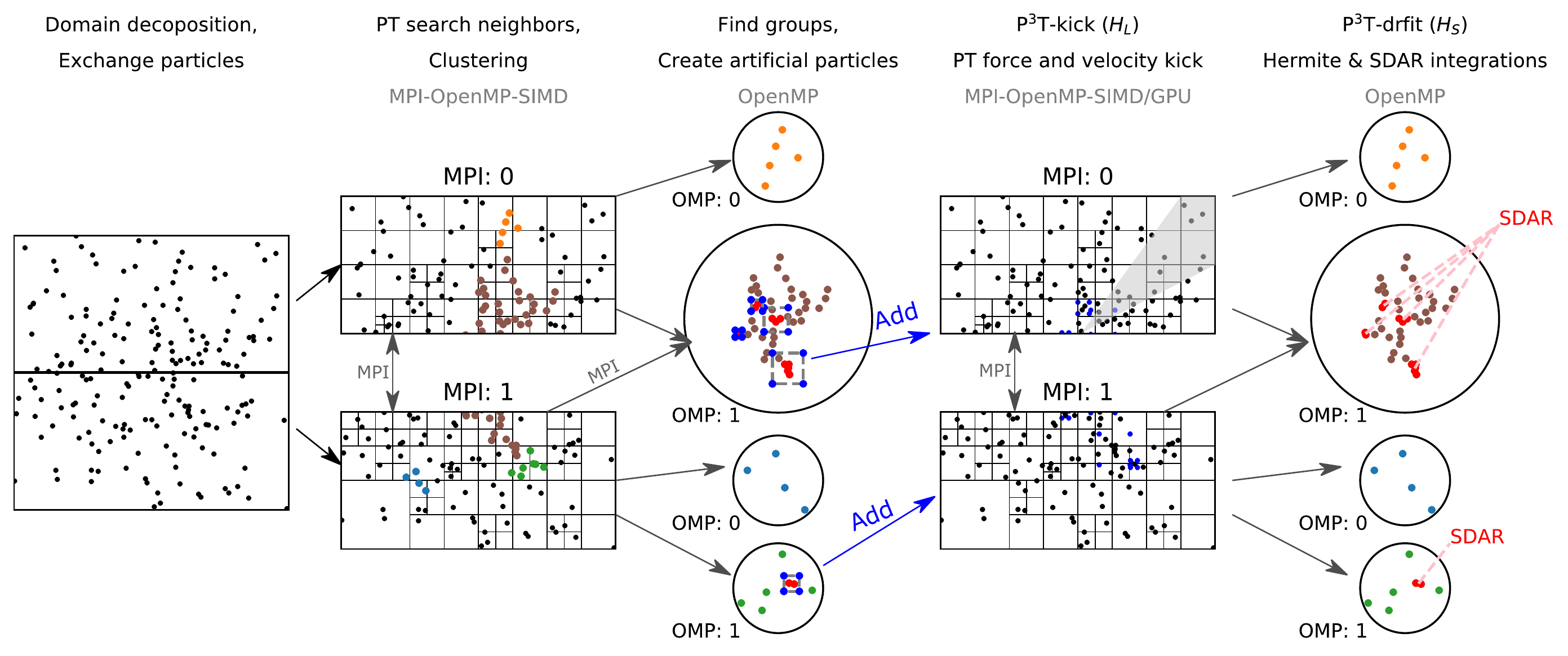}
  \caption{The schedule of one kick-drift-kick cycle of \petar. In order to have a clear view, the two-dimensional particle system is illustrated here. From left to right:
    1) Domain decomposition splits particles in space and distributes them to different MPI processes. This step is done every few cycles.
    2) Particle-tree (PT) is constructed for searching neighbours and clustering. Individual clusters are marked as different colours.
    3) For individual clusters, groups of multiple systems (binaries, triples ...; marked as red points) are detected and artificial particles (tidal-tensor and orbit-sampling / pseudo- particles; blue points) are added to particle systems. Artificial particles are represented by four points along a rectangle.
    4) PT is constructed with artificial particles and used to calculate the long-range force. Then velocities of particles are kicked.
    5) For individual clusters, particle positions and velocities are integrated by the Hermite and SDAR methods (drift).
  }
  \label{fig:petar}
\end{figure*}

We introduce our new hybrid $N$-body code, \petar, which combines the \pppt~method \citep{Oshino2011,Iwasawa2015,Iwasawa2016,Iwasawa2017} and the slow-down time-transformed symplectic integrator \citep[SDAR;][]{Wang2020}. 
The framework of \textsc{pentacle} is the base of the code \citep{Iwasawa2017}.
The parallelization framework for developing particle simulation codes \citep[\fdps;][]{Iwasawa2016,Iwasawa2020} are used to deal with the particle-tree construction and long-range force calculation.
The \sdar code, which combines the fourth-order Hermite and the SDAR integrators\footnote{Notice that \sdar and SDAR (with different font styles) are the names of the code and the algorithm, respectively.}. is used for the short-range interaction.
Fig.~\ref{fig:petar} show how \petar~works for one cycle of integration.
It can be summaries as:
\begin{enumerate}
\item {\bf Decompose domains}: distribute particles to different MPI processes.
\item {\bf Search neighbours and clustering}: construct particle tree, search neighbours for each particle and gather all nearby particles into individual clusters (Section~\ref{sec:clustering}).
\item {\bf Find groups and create artificial particles}: in each cluster, find sub-systems (groups) and if necessary, create artificial particles for each group (Section~\ref{sec:sdar} and \ref{sec:tt}).
\item {\bf Calculate long-range force and kick velocities (\pkick)}: construct particle tree that includes artificial particles, calculate the long-range interaction and kick the velocities of all particles.
\item {\bf Integrate motions in each clusters (\pdrift)}: in each cluster, use the Hermite and SDAR method to integrate the motions controlled by the short-range interaction.
\end{enumerate}
The kick-drift-kick mode (Eq.~\ref{eq:lfmap}) is used, thus in the final step, (i)-(iv) are executed once more.
The first and the last \pkick~take the half of $\dts$.

\subsection{Mass-dependent changeover function}
\label{sec:changeover}

In \textsc{pentacle}, the seventh-order polynomial type of changeover function $K(\rij)$ (the derivative of $W(\rij)$ with respect to $\rij$) is implemented \citep{Iwasawa2017}.
This $K(\rij)$ ensures that all terms of derivatives of force used in the fourth-order Hermite integrator have a smooth changes at the boundary of the changeover range.
However, there are two limits.
Firstly, the changeover function for potential $W(\rij)$ contains the term of $\log(\rij)$, which is computational expensive.
Secondly, the changeover range is fixed for all particles, i.e., $\rin$ and $\rout$ are constant.
However, in star clusters, the mass spectrum has a wide range where the ratio between the maximum mass and minimum mass can be as large as $10^4$.
If a very massive object like super massive black hole exists, the ratio can be $10^8$.
For the same distance, the force from massive objects are larger than that of the low-mass objects.
Thus the fixed changeover range cannot properly handle the systems with a wide mass spectrum.

We introduce a mass-dependent changeover function, where each particle has an individual changeover range ($\rini$, $\routi$).
The cubic root of the particle mass is used as the coefficient to determine the boundary:
\begin{equation}
  \begin{aligned}
    \rini = & \mathrm{max} \left(1, \frac{m_i}{\mave} \right )^{\frac{1}{3}}\rinr, \\
    \routi =& \mathrm{max} \left(1, \frac{m_i}{\mave} \right )^{\frac{1}{3}}\routr,\\
  \end{aligned}
  \label{eq:rcut}
\end{equation}
where $\rinr$ and $\routr$ are the reference of a fixed changeover range, and $\mave$ is the average mass of the system.
The minimum mass factor is $1.0$ so that low-mass particles can avoid too small changeover radii.

If two particles $i$ and $j$ has a separation $\rij<\rinij$, the perturbation from a distant particle $k$ vs. the internal force between the two particles can be estimated as 
\begin{equation}
  \fr(\rcmk) = \frac{\mk}{\mi+\mj} \left( \frac{\rij}{\rcmk} \right )^3,
  \label{eq:fp}
\end{equation}
where $\rcmk$ is the distance between the centre-of-the-mass of the pair $i$ and $j$ and the perturber $k$.
If $\mk>\mi+\mj$, the changeover radii between the pair and the perturber are determined by $\mk$.
Eq.~\ref{eq:fp} indicates that
\begin{equation}
  \fr(\rink) = \mathrm{min} \left (1, \frac{\mave}{\mi+\mj} \right )\left ( \frac{\rij}{\rinr} \right )^3.
\end{equation}
Thus, $\fr$ at the changeover boundary is independent of $\mk$.
Therefore, Eq.~\ref{eq:rcut} is sufficient to handle the tidal perturbation from massive objects.

\begin{figure}
  \centering
  \includegraphics[width=0.6\columnwidth]{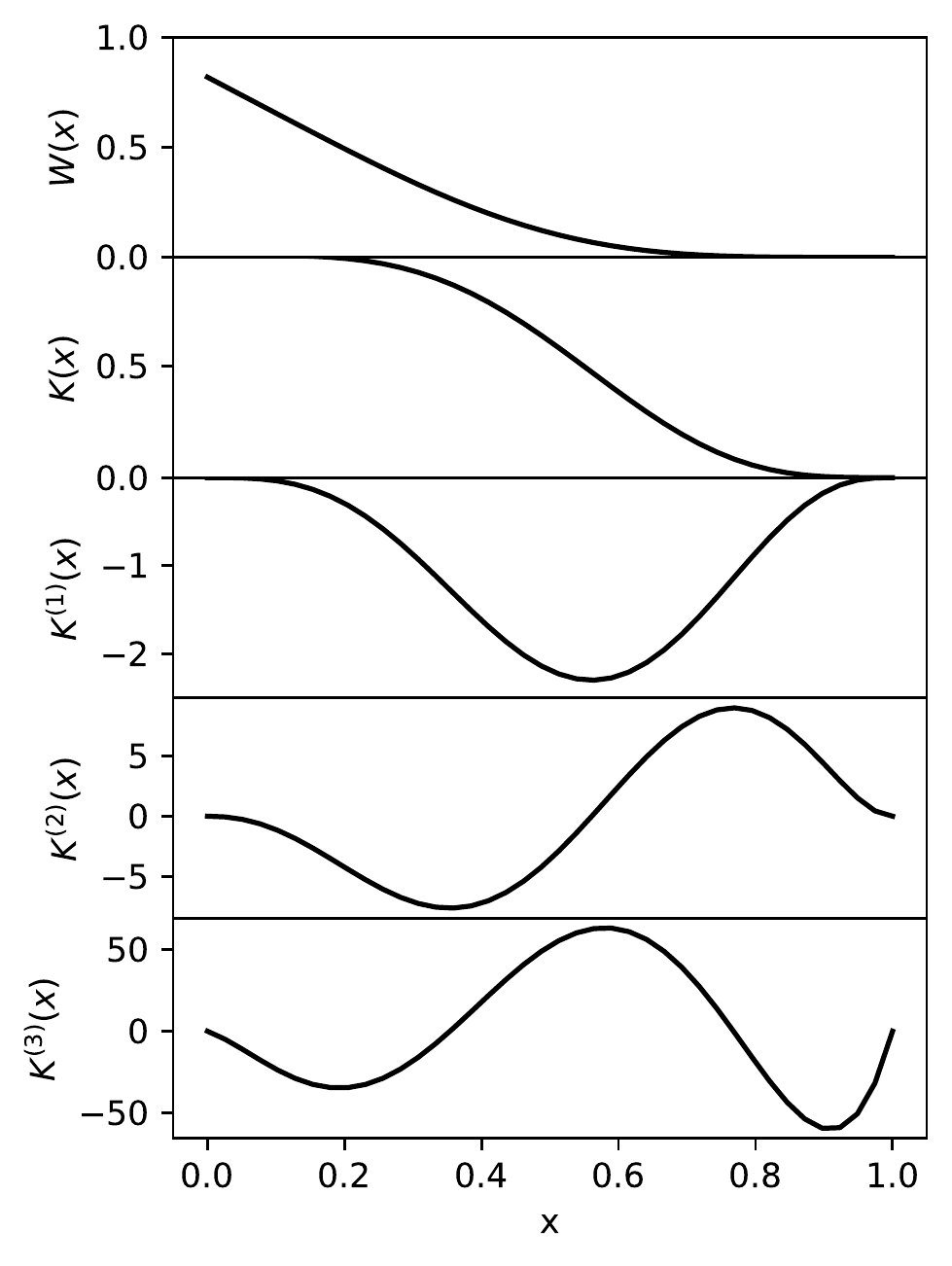}
  \caption{The shapes of the changeover functions for potential, $W(x)$, and for force, $\WF(x)$, and its $n$-th time derivatives, $\WF^{(n)}(x)$. Here $\rinij=1$ and $\routij=10$.}
  \label{fig:cof}
\end{figure}

On the other hand, to avoid logarithmic function, we use the eighth-order polynomial function as the changeover function for potential: 
\begin{equation}
  W(x)  = 
  \begin{cases}
    \Lambda (1-2 x) & (x\le 0) \\
    \Lambda (1-2 x) - 1 + f(x) & (0 < x < 1)\\
    0  & (x\ge 1) \\
  \end{cases} 
\end{equation}
where 
\begin{equation}
  \begin{aligned}
    \WU(x) & = 1 + \Lambda x^5 \left( 14 - 28 x + 20 x^2 -5 x^3 \right) \\
    x & = \frac{\rij-\rinij}{\routij-\rinij} \\
    \Lambda & = \frac{\routij-\rinij}{\routij+\rinij} \\
    \rinij &= \max(\rini,\rinj) \\
    \routij &= \max(\routi, \routj).\\
  \end{aligned} 
\end{equation}
We use $x$ instead of $\rij$ in the formulas.
The changeover function for force has the form:
\begin{equation}
  \WF(x) =
  \begin{cases}
    1 & (x \le 0) \\
    (x - 1)^4\times  & \\
    \quad \left( 1 + 4x + 10 x^2 + 20 x^3 + 35 \Lambda x^4  \right) & (0 < x < 1)\\
    0 & (x \ge 1) \\
  \end{cases}
  \label{eq:Wij}
\end{equation}
For $0<x<1$, $W(x)$ and $\WF(x)$ are related to $\WU(x)$ by 
\begin{equation}
  \begin{aligned}
    W(x) & = \WU(x) - \WU(1) \frac{\rij}{\routij} \\
    \WF(x) &= \WU(x) - \left (x + \frac{\rinij}{\routij-\rinij}\right) \WU^{(1)}(x), \\
  \end{aligned}
\end{equation}
where the number in superscript ``$()$'' indicates the times of derivative with respect to $x$.
The second term in the expression of $W(x)$ is an offset ensuring that the potential becomes zero at $\routij$.
At the boundary ($x=0,1$), the changeover functions have values:
\begin{equation}
  \begin{aligned}
    W(1)        =&0,   &              &  \\
    \WF(0)      =&1,   & \WF(1)      =&0, \\ 
    \WF^{(1)}(0) =& 0, & \WF^{(1)}(1)=&0, \\
    \WF^{(2)}(0) =& 0, & \WF^{(2)}(1)=&0, \\
    \WF^{(3)}(0) =& 0, & \WF^{(3)}(1)=&0.\\
  \end{aligned}
  \label{eq:cof}
\end{equation}
The potential and force of $\HS$ reduces to zero after $x\ge1$.
All derivatives of $\WF(x)$ are zero at the boundary.
These ensure that the higher-order (up to $3$) derivatives of force used in the fourth-order Hermite integrator have smooth curves at $\rinij$ and $\routij$.
Fig.~\ref{fig:cof} show the examples of these functions with $\routij=10$ and $\rinij=1$.

Since $\rinij$ and $\routij$ take the maximum values from the changeover radii of $i$ and $j$ particles,
the strong force from the massive particles are always preferentially included in $\HS$ for a high accuracy.
Besides, for both particles, the changeover range is identical, thus the force is symmetric.

\subsubsection{Varying changeover radii}

To ensure that the \pppt~method is symplectic, once the changeover radii of all particles are determined, they should keep unchanged during the integration.
However, in real star clusters, masses of stars evolve due to the stellar-wind driven mass loss or the mass transfer and mergers of two stars.
Besides, binary can form, disrupt and change members.
Thus, after a certain time, the changeover radii of one star or binary may not be suitable anymore and need to be recalculated by using the new mass.
This breaks the symplectic properties of the integrator.
To minimize the effect, the modification of changeover radii and masses can only be done after a complete leapfrog step.



\subsection{Slow-down time-transformed symplectic integrator}
\label{sec:sdar}

Short-period binaries are challenging not only due to the time consuming integration, but also because of the large cumulative numerical errors after many orbits.
The symplectic integrator can conserve Hamiltonian and angular momentum for the long-term evolution.
However, it requires a constant integration step, thus very small time steps have to be used for highly eccentric Kepler orbits and close encounters.
One way to avoid small steps is to use the extended phase-space Hamiltonian,
\begin{equation}
  \Gamma(\Wv) = g(\Wv) \left[H(\wv,t) - H(\wvz,0)\right],
  \label{eq:gamma}
\end{equation}
where $H(\wv,t)$ is the standard Hamiltonian, $g(\Wv)$ is time-transformation function and $\Wv$ is the extended phase-space vector that contains $\wv$ and new pair of the coordinate, $t$, and the corresponding conjugate momentum $\pt$.
By introducing the new differential variable, $s$, the equation of motion can be described as
\begin{equation}
  \frac{\diff \Wv}{\diff s} = \{ \Wv, \Gamma(\Wv) \}
\end{equation}
Thus, the time step and integration step are decoupled via the time transformation function.
For eccentric orbits, time steps can vary based on the requirement of accuracy and efficiency while $\diff s$ keeps constant.
In order to use explicit symplectic method, $g(\Wv)$ should be designed to make $\Gamma(\Wv)$ separable like Eq.~\ref{eq:split}.
\cite{Mikkola1999} and \cite{Preto1999} provided such a solution by using
\begin{equation}
  g(\Wv) = \frac{f(T(\Pv)) - f(-U(\Rv))}{T(\Pv) + U(\Rv)} ,
  \label{eq:g}
\end{equation}
where $T(\Pv)$ and $U(\Rv)$ are kinetic and potential energy in the extended phase space.
When $f(x) = \log(x)$ with a leapfrog integrator in the drift-kick-drift mode, the Kepler orbit can be integrated very accurately with only round-off errors in positions and velocities and a phase error of time.
In \cite{Mikkola1999}, this is named ``Algorithmic regularization (AR)''.
The AR method can well solve the issue of the long-term cumulative errors.

On the other hand, the slow-down method described in Section~\ref{sec:slowdown} can reduce the total integration steps for weakly perturbed binaries.
\cite{Wang2020} combined the slow-down and AR methods and developed the SDAR algorithm to efficiently and accurately integrate the few-body systems.
In their work, $\kappa$ is calculated by the perturbation criterion and timescale criterion.
We set the tree time step, $\Cs \dts$, as the maximum timescale criterion where $\Cs$ is a coefficient larger than one.
In such case, if $\kappa$ of a weakly perturbed binary reaches the maximum value, $\dts$ is small enough to resolve the orbit of the binary in order to provide the correct \pkick.
With the slow-down method, the actual integration steps of binaries for a given physical time interval are decoupled from the real $\tbin$ but depends on the perturbation and $\dts$.
Since most short-period binaries are weakly perturbed in a star cluster and their $\tbin \ll \dts$, the total number of integration steps are significantly reduced by a few thousand times.
Thus, the SDAR method is the major algorithm in the \petar~code to solve the multiple timescale issue.

In the \pdrift~step, we use the SDAR method for compact groups of particles and the Hermite integrator for integrating the motions of singles and centre-of-the-mass of groups.
The particle groups are determined by a distant criterion, $\rij<\rgij$, where
\begin{equation}
  \rgij = \theta \rinij,
  \label{eq:rbin}
\end{equation}
and $\theta$ is the opening angle of the PT method.
This is not a strict criterion.
We use Eq.~\ref{eq:rbin} so that any pair of members in a group are always inside their inner boundaries of changeover ranges.
Thus, the SDAR method only need to deal with the Newtonian force.
This avoids the complexity of using the SDAR method and changeover functions together.
Besides, if one particle receives a long-range interaction from the group, its members are inside the angle  $\theta$ viewing from this particle.


\subsection{Clustering}
\label{sec:clustering}

In Section~\ref{sec:bottleneck}, we show that the switch between the regularization method and the Hermite integrator is expensive with an $O(N\langle \Nb \rangle)$ memory access and an $O(N)$ force calculation in \textsc{nbody6(++gpu)}.
This issue is general for the hybrid methods that use neighbour list and need force calculation for the centre-of-the-masses of groups.
In \petar, we use the clustering scheme to avoid such expensive switching.
This scheme is originally implemented in the \textsc{pentacle} code \citep{Iwasawa2017} with an uniform neighbour radius.
Here we describe the idea and introduce the improved algorithm based on the orbit-dependent neighbour criterion.

After searching short-range interacting neighbours and before Hermite integration, particles are collected together into different clusters (Fig.~\ref{fig:petar}).
The clustering scheme ensures that any member in one cluster have all its neighbours inside the same cluster.
In such case, particles outside this cluster only provide the long-range interaction to the members.
Thus, during the integration of the short-range interactions (\pdrift), each cluster is isolated to others and can be integrated in parallel.
This feature leads to a great advantage: the switch between the different integration methods in one cluster only affect the neighbour lists and forces of the local members.
Since the typical number of members per cluster is a small fraction of the total number of particles, the computational cost of switching is much less.

On the other hand, when MPI parallelization is used, sometimes one cluster may contain members crossing multiple MPI processes (like the cluster, OMP:1, shown in Fig.~\ref{fig:petar}).
In such case, one MPI process is chosen to be the host for the cluster and others send particle data to it.

\subsubsection{Orbit-dependent neighbour criterion}

The number of members in clusters determine the performance of \pdrift.
Thus, it is important to choose a proper neighbour searching criterion.
In star clusters with a mass spectrum, we cannot apply the uniform neighbour radius as in \textsc{pentacle}.
Instead, we determine the individual neighbour searching radius, $\rnbi$, based on $\routi$ and the velocity.
Firstly, for each particle, $\rnbi$ must be longer than $\routi$.
However, we cannot set these two radii the same because during \pdrift, particles that are initially not inside the short-range interaction region can move closer and penetrate the boundary.
Therefore, $\rnbi$ should be long enough to capture such potential neighbours.
One safe way is to include the velocity information that
\begin{equation}
  \rnbi = \routi + \nbfac |\vi| \dts,
  \label{eq:rnb}
\end{equation}
where $\vi$ is the particle velocity and $\nbfac$ is a free coefficient (we use $3.0$ for safety).

However, this criterion is independent on the direction of velocity.
If the particle velocity is large, $\rnbi$ is significantly long that a huge cluster can form.
Unfortunately, high-velocity particles are commonly generated via few-body interactions in star clusters.
Thus, we need to reduce the neighbour numbers for these particles.
For a high-velocity particle, only neighbours along its path are important and most particles inside $\rnbi$ are not real neighbours.
To avoid including these unnecessary neighbours, a three-dimensional neighbour searching criterion which depends on the direction of velocity is needed.
However, such criterion is not computationally efficient and is not supported by the current version of \fdps.
To solve this issue, we apply a two-stage method:
\begin{enumerate}
\item Obtain the neighbour candidates by applying the spherically symmetric neighbour searching using Eq.~\ref{eq:rnb}. 
\item Select true neighbours if the candidate $j$ has the Kepler orbital pericentre separation, $\rpij<\nbvfac \rnbi$, where $\nbvfac$ is a free coefficient (e.g. $1.5$).
\end{enumerate}
The first step has the calculation cost of $O(N\log N)$ by using the particle-tree method.
Since a proper $\dts$ leads to a large fraction of particles with no neighbour candidates, the cost of evaluations of Kepler orbital pericentre at the second step is not expensive.
Thus, this method is efficient to deal with the problem of high-velocity particles.

\subsection{Artificial particle algorithm for weak perturbation}
\label{sec:tt}

Both the changeover function and $\dts$ influence the performance and accuracy of the simulations.
We can understand this by analysing a situation where a binary receives the long-range perturbation force.
Although the long-range force is much weaker compared to the internal force of the binary, it is still important to ensure that $\dts<\tbin$.
Otherwise a random phase of binary is chosen to evaluate the long-range force, which does not represent the correct perturbation.
This is the same for the counter force.
If the binary is very massive, this error can be significant.

However, keeping $\dts<\tbin$ is difficult since $\tbin$ can be very small.
In Section~\ref{sec:sdar}, we show that the slow-down method can artificially increase $\tbin$, which helps to avoid too small $\dts$.
But only weakly perturbed binaries can have large enough $\kappa$.
When a tight binary has close neighbours, the effective $\tbin$ can be much smaller than $\dts$.
This can frequently happen in star clusters.

To solve this issue, we introduce the ``artificial particle algorithm''.
In this algorithm, instead of calculating the long-range force once and giving a large velocity kick per $\dts$, we can construct the local potential (tidal-tensor) field near the binary and use it to calculate a smooth evolution of the long-range perturbation every AR step.
The tidal-tensor field can be obtained by measuring the long-range forces of a group of artificial particles near the binary.
On the other hand, another group of artificial particles along or near the orbit of the binary can be used to to represent the correct orbit-averaged counter-force.

This algorithm increases the total number of particles, thus the number of long-range interactions becomes more.
However, this additional cost can be easily reduced by increasing the number of computing cores.
If we use small $\dts$, there is no such simple solution.
Besides, adding artificial particles is easy to implement.
This is especially convenient for using the \fdps~library and the accelerators such as SIMD and GPU.

\subsubsection{Tidal-tensor}

Here we describe the algorithm to obtain the local tidal-tensor field near the binary.
Based on a 3-dimension Taylor expansion, the acceleration of a particle at an arbitrary position near a fixed centre can be evaluated by:
\begin{equation}
  \begin{split}
    \mathbf{A}_i(\mathbf{r}) =& \mathbf{A}_i(\mathbf{r}_{\mathrm{cm}})  + \mathbf{A}_{ij} \cdot (\mathbf{r}-\mathbf{r}_{\mathrm{cm}})_{j} \\ & + \frac{1}{2!} (\mathbf{r}-\mathbf{r}_{\mathrm{cm}})_{j}^T \cdot \mathbf{A}_{ijk} \cdot (\mathbf{r}-\mathbf{r}_{\mathrm{cm}})_{k} + ... \\
    A_{ij} =& \frac{\partial A_{i}}{\partial r_{j}}\\
    A_{ijk} =& \frac{\partial^2 A_{i}}{\partial r_{j}\partial r_{k}}
    \end{split}
  \label{eq:tt}
\end{equation}
where $A_{ij}$ and $A_{ijk}$ are individual components of the tensors, $\mathbf{A}_{ij}$ and $\mathbf{A}_{ijk}$, respectively; and ``$\cdot$'' represents matrix multiplication.
In the $\mathrm{P^3T}$ method, the numerical long-range forces are constant within one $\dts$, so should be the tensor field.
Thus, we only need to measure the tensors once per $\dts$.
Then, using Eq.~\ref{eq:tt}, the long-range perturbation on an arbitrary orbital phase of the binary can be evaluated during the \pdrift.
For the gravitational field, $\mathbf{A}_{ij}$ and $\mathbf{A}_{ijk}$ are symmetric tensors.
The number of elements of the first three orders are $3$ ($\mathbf{A}_i(\mathbf{r}_{\mathrm{cm}})$), $6$ ($\mathbf{A}_{ij}$) and  $10$ ($\mathbf{A}_{ijk}$), respectively.
Thus, the second-order method has totally $9$ elements and the third-order has $19$.

\begin{figure}
  \centering
  \includegraphics[width=0.7\columnwidth]{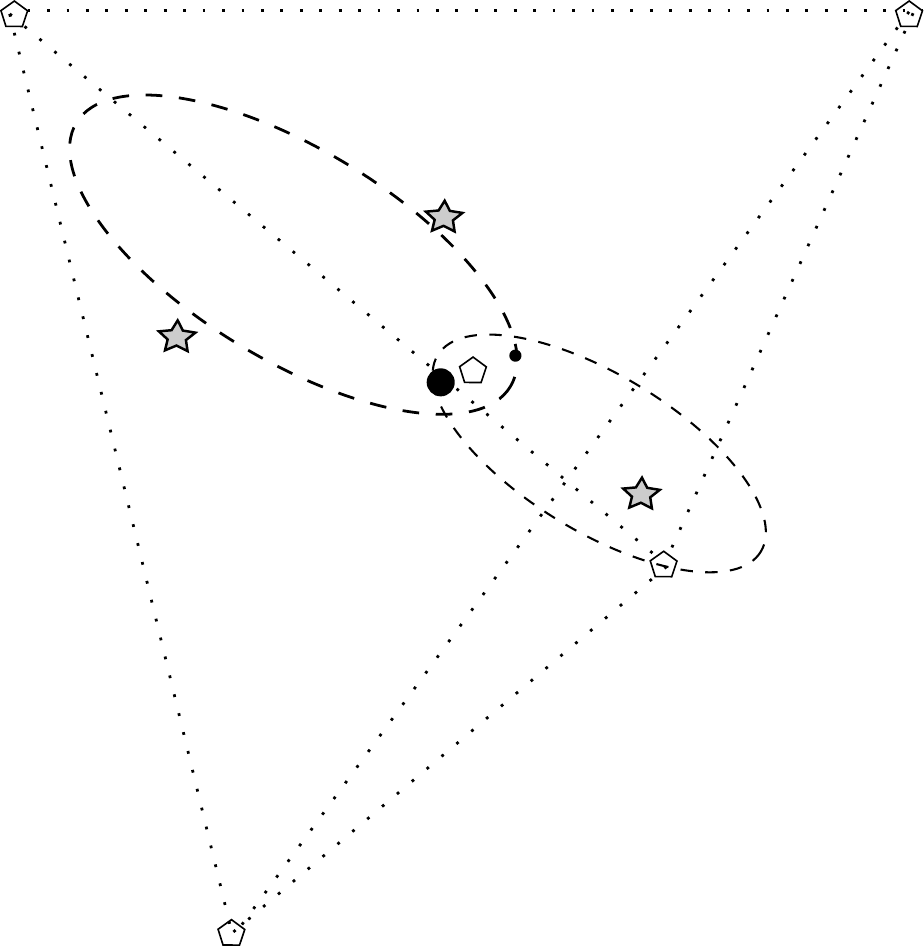}\\
  $2^{\mathrm{nd}}$-order tidal-tensor and pseudoparticle multipole\\ 
  \includegraphics[width=0.7\columnwidth]{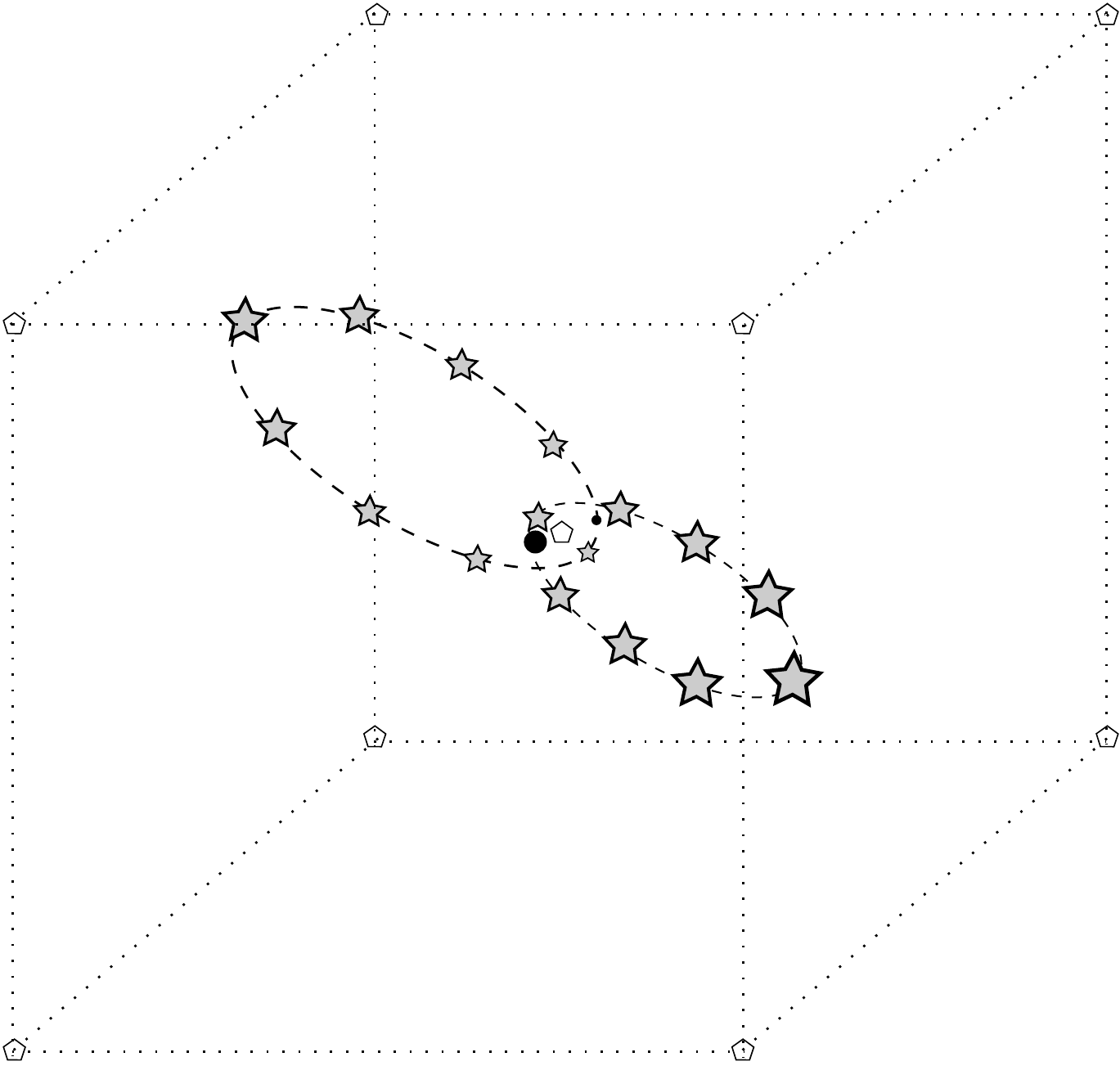} \\
  $3^{\mathrm{rd}}$-order tidal-tensor and orbit-sampling \\
  \caption{The illustration of the spatial distribution of the artificial particles in the tidal-tensor, orbit-sampling and pseudoparticle multipole methods.
    Open pentagons at the corners and the centre are tidal-tensor measure points.
    Filled stars along the orbits of two binary components are pseudoparticles or orbit-sampling particles used for evaluating the orbit-averaged long-range counter-force to distant particles.
    Notice here two combinations are shown, it is possible to combine different orders of tidal tensors with either pseudoparticle multipole or orbit-sampling.
  }
  \label{fig:tt}
\end{figure}

To obtain these tensor elements, we can create measure points (zero-mass artificial particles) near the centre-of-the-mass of the binary.
These artificial particles obtain the long-range interactions during \pkick.
Using the three-dimensional accelerations of one measure point in Eq.~\ref{eq:tt}, we can get three independent linear equations of the tensor elements.
At least $3$ measure points are needed to obtain the unique values of tensors up to the second order.
The third-order case requires $7$ points.

The acceleration of the centre-of-the-mass can be used to directly measure the zero-order acceleration, $\mathbf{A}(\mathbf{r}_{\mathrm{cm}})$.
We collect other components of the tensors in one-dimensional vectors for the second- ($2^{\mathrm{nd}}$) and third- ($3^{\mathrm{rd}}$) order methods:
\begin{equation}
  \centering
  \begin{matrix}
    \mathbf{T}_{i} (2^{\mathrm{nd}}) = \left[ \right.  &  A_{xx} & A_{xy}  & A_{xz}  & A_{yy}  &  A_{yz} & A_{zz} & \left. \right]\\
    \mathbf{T}_{i} (3^{\mathrm{rd}}) = \left[ \right. &  A_{xx} & A_{xy}  & A_{xz}  & A_{yy}  &  A_{yz} & A_{zz} & \\
                                   &  A_{xxx} & A_{xxy} &  A_{xxz} & A_{xyy} & A_{xyz} & A_{xzz}& \\
                                   &  A_{yyy} & A_{yyz} & A_{yzz} & A_{zzz} &      &        &\left. \right].\\
  \end{matrix}
\end{equation}
The accelerations of measure points excluding $\mathbf{A}(\mathbf{r}_{\mathrm{cm}})$ can be also collected as an one-dimensional vector:
\begin{equation}
  \begin{matrix}
    \mathbf{A'}_{\mathrm j} =
      & \left[ \right. A'_{\mathrm{x}}(\mathbf{r}_{\mathrm 1}) & A'_{\mathrm{y}}(\mathbf{r}_{\mathrm 1}) & A'_{\mathrm{z}}(\mathbf{r}_{\mathrm 1})\\
      &  A'_{\mathrm{x}}(\mathbf{r}_{\mathrm 2}) & A'_{\mathrm{y}}(\mathbf{r}_{\mathrm 2}) & A'_{\mathrm{z}}(\mathbf{r}_{\mathrm 2})\\
      & ... && \left. \right],\\
  \end{matrix}
\end{equation}
where $\mathbf{A}'(\mathbf{r})=\mathbf{A}(\mathbf{r})-\mathbf{A}(\mathbf{r}_{\mathrm{cm}})$ and the suffixes, $1,2,3,...$, are the indices of points.

Based on Eq.~\ref{eq:tt}, $\mathbf{T_{\mathrm i}}$ and $\mathbf{A}_{\mathrm j}$ can be described by a linear mapping:
\begin{equation}
  \mathbf{M}_{ij} \mathbf{T}_{\mathrm i} = \mathbf{A}_{\mathrm j}
  \label{eq:map}
\end{equation}
Once the generalized inverse matrix, $\mathbf{M}_{ij}^{-1}$, is obtained, $\mathbf{T}_{\mathrm i}$ can be easily calculated once $\mathbf{A}_{\mathrm j}$ are measured.

In principle $2$ points excluding the centre-of-the-mass are enough for the second-order method.
However, we can only obtain the two-dimensional information in a plane.
Thus we use $4$ points locating at the corners of a regular tetrahedron (see the upper panel in Fig.~\ref{fig:tt}).
In the third-order case, the corners of a regular octahedron can provide $6$ points.
However, in such case, we find the rank of $\mathbf{M}_{ij}$ is not full so that $\mathbf{M}_{ij}^{-1}$ cannot be constructed.
Thus, we use $8$ points locating at the corners of a cube instead (the lower panel in Fig.~\ref{fig:tt}).
Although two additional points are needed for both two methods, we obtain the benefit that the condition numbers of the matrices (the maximum singular value vs. the minimum) are small: $2$ and $12.7$ for the second- and the third-order methods, respectively.
This means that the relative error of measurement inherited from $\mathbf{A}_{\mathrm j}$ can be maximally enlarged by a factor of $2$ or $12.7$ in $\mathbf{M}_{ij}^{-1}$.
The exact values of the elements in $\mathbf{M}_{ij}^{-1}$ can also be obtained easily.
Table~\ref{tab:A} provides the complete formulas to evaluate $\mathbf{T}_{\mathrm i}$ and to calculate the acceleration at any $\mathbf{r}$.
The corresponding coordinates of the measure points are also provided.

  
\begin{table*}
  \centering
  \caption{The second- (upper block) and third-order (lower block) tidal tensor methods. Each block contains three parts: (1) the formulas to calculate the tensor coefficients, $\mathbf{A}_{ij} $ and $\mathbf{A}_{ijk}$, where the centre-of-the-mass acceleration is subtracted in $\mathbf{A}'(\mathbf{r}_{\mathrm i})$.
    (2) the formulas to calculate the acceleration $\mathbf{A}(\mathbf{r})$ for a given position.
    where $\mathbf{r}'$ ($x'$, $y'$, $z'$) is the coordinate referring to the centre-of-the-mass ($\mathbf{r}-\mathbf{r}_{\mathrm{cm}}$).
    (3) the coordinates of the measure points, where $d_{\mathrm c}$ is the half length of the edge of the tetrahedron or the cube.
  }
  \label{tab:A}
  \begin{tabular}{llllllllll}
    \hline\hline
    \multicolumn{10}{l}{Second-order tidal tensor with 4 measure points at corners of a regular tetrahedron}\\
    \hline
    $A_{xx}$ & $ = $ & \multicolumn{8}{l}{$ \frac{1}{d_{\mathrm c}} \left [ \right. + \frac{1}{2} A'_{x}(\mathbf{r}_1)- \frac{1}{2} A'_{x}(\mathbf{r}_2)\left. \right ]$}\\
$A_{xy}$ & $ = $ & \multicolumn{8}{l}{$ \frac{1}{d_{\mathrm c}} \left [ \right. + \frac{1}{4} A'_{y}(\mathbf{r}_1)- \frac{1}{4} A'_{y}(\mathbf{r}_2)+ \frac{1}{4} A'_{x}(\mathbf{r}_3)- \frac{1}{4} A'_{x}(\mathbf{r}_4)\left. \right ]$}\\
$A_{xz}$ & $ = $ & \multicolumn{8}{l}{$ \frac{1}{d_{\mathrm c}} \left [ \right. - \frac{\sqrt{2}}{8} A'_{x}(\mathbf{r}_1)+ \frac{1}{4} A'_{z}(\mathbf{r}_1)- \frac{\sqrt{2}}{8} A'_{x}(\mathbf{r}_2)- \frac{1}{4} A'_{z}(\mathbf{r}_2)+ \frac{\sqrt{2}}{8} A'_{x}(\mathbf{r}_3)+ \frac{\sqrt{2}}{8} A'_{x}(\mathbf{r}_4)\left. \right ]$}\\
$A_{yy}$ & $ = $ & \multicolumn{8}{l}{$ \frac{1}{d_{\mathrm c}} \left [ \right. + \frac{1}{2} A'_{y}(\mathbf{r}_3)- \frac{1}{2} A'_{y}(\mathbf{r}_4)\left. \right ]$}\\
$A_{yz}$ & $ = $ & \multicolumn{8}{l}{$ \frac{1}{d_{\mathrm c}} \left [ \right. - \frac{\sqrt{2}}{8} A'_{y}(\mathbf{r}_1)- \frac{\sqrt{2}}{8} A'_{y}(\mathbf{r}_2)+ \frac{\sqrt{2}}{8} A'_{y}(\mathbf{r}_3)+ \frac{1}{4} A'_{z}(\mathbf{r}_3)+ \frac{\sqrt{2}}{8} A'_{y}(\mathbf{r}_4)- \frac{1}{4} A'_{z}(\mathbf{r}_4)\left. \right ]$}\\
$A_{zz}$ & $ = $ & \multicolumn{8}{l}{$ \frac{1}{d_{\mathrm c}} \left [ \right. - \frac{\sqrt{2}}{4} A'_{z}(\mathbf{r}_1)- \frac{\sqrt{2}}{4} A'_{z}(\mathbf{r}_2)+ \frac{\sqrt{2}}{4} A'_{z}(\mathbf{r}_3)+ \frac{\sqrt{2}}{4} A'_{z}(\mathbf{r}_4)\left. \right ]$}\\
    \hline
    $A_{x}(\bm{r}) $ & $ = $ & \multicolumn{8}{l}{$ A_{x}(\mathbf{r}_{\mathrm{cm}})  + A_{xx}x' + A_{xy}y' + A_{xz}z' $}\\
    $A_{y}(\bm{r}) $ & $ = $ & \multicolumn{8}{l}{$ A_{y}(\mathbf{r}_{\mathrm{cm}})  + A_{xy}x' + A_{yy}y' + A_{yz}z' $}\\
    $A_{z}(\bm{r}) $ & $ = $ & \multicolumn{8}{l}{$ A_{z}(\mathbf{r}_{\mathrm{cm}})  + A_{xz}x' + A_{yz}y' + A_{zz}z' $}\\
    \hline
    $\mathbf{r}'_{\mathrm i}/d_{\mathrm c}$  & $=$ & $\left[\begin{matrix} 1 & 0 & - \frac{\sqrt{2}}{2} \end{matrix}\right]$ &$\left[\begin{matrix} -1 & 0 & - \frac{\sqrt{2}}{2} \end{matrix}\right]$ &$\left[\begin{matrix} 0 & 1 & \frac{\sqrt{2}}{2} \end{matrix}\right]$ &$\left[\begin{matrix} 0 & -1 & \frac{\sqrt{2}}{2} \end{matrix}\right]$ &&&&\\
    \hline\hline
    \multicolumn{10}{l}{Third-order tidal tensor with 8 measure points at corners of a cube}\\
    \hline
$A_{xx}$ & $ = $ & \multicolumn{8}{l}{$ \frac{1}{d_{\mathrm c}} \left [ \right. + \frac{1}{4} A'_{x}(\mathbf{r}_1)- \frac{1}{4} A'_{x}(\mathbf{r}_3)+ \frac{1}{4} A'_{x}(\mathbf{r}_5)- \frac{1}{4} A'_{x}(\mathbf{r}_7)\left. \right ]$}\\
$A_{xy}$ & $ = $ & \multicolumn{8}{l}{$ \frac{1}{d_{\mathrm c}} \left [ \right. + \frac{1}{8} A'_{y}(\mathbf{r}_1)+ \frac{1}{8} A'_{x}(\mathbf{r}_2)- \frac{1}{8} A'_{y}(\mathbf{r}_3)- \frac{1}{8} A'_{x}(\mathbf{r}_4)+ \frac{1}{8} A'_{y}(\mathbf{r}_5)+ \frac{1}{8} A'_{x}(\mathbf{r}_6)- \frac{1}{8} A'_{y}(\mathbf{r}_7)- \frac{1}{8} A'_{x}(\mathbf{r}_8)\left. \right ]$}\\
$A_{xz}$ & $ = $ & \multicolumn{8}{l}{$ \frac{1}{d_{\mathrm c}} \left [ \right. - \frac{1}{12} A'_{x}(\mathbf{r}_1)+ \frac{1}{12} A'_{z}(\mathbf{r}_1)- \frac{1}{12} A'_{x}(\mathbf{r}_2)- \frac{1}{12} A'_{x}(\mathbf{r}_3)- \frac{1}{12} A'_{z}(\mathbf{r}_3)- \frac{1}{12} A'_{x}(\mathbf{r}_4)+ \frac{1}{12} A'_{x}(\mathbf{r}_5)+ \frac{1}{12} A'_{z}(\mathbf{r}_5)+ \frac{1}{12} A'_{x}(\mathbf{r}_6)$}\\ & & \multicolumn{8}{l}{$+ \frac{1}{12} A'_{x}(\mathbf{r}_7)- \frac{1}{12} A'_{z}(\mathbf{r}_7)+ \frac{1}{12} A'_{x}(\mathbf{r}_8)\left. \right ]$}\\
$A_{yy}$ & $ = $ & \multicolumn{8}{l}{$ \frac{1}{d_{\mathrm c}} \left [ \right. + \frac{1}{4} A'_{y}(\mathbf{r}_2)- \frac{1}{4} A'_{y}(\mathbf{r}_4)+ \frac{1}{4} A'_{y}(\mathbf{r}_6)- \frac{1}{4} A'_{y}(\mathbf{r}_8)\left. \right ]$}\\
$A_{yz}$ & $ = $ & \multicolumn{8}{l}{$ \frac{1}{d_{\mathrm c}} \left [ \right. - \frac{1}{12} A'_{y}(\mathbf{r}_1)- \frac{1}{12} A'_{y}(\mathbf{r}_2)+ \frac{1}{12} A'_{z}(\mathbf{r}_2)- \frac{1}{12} A'_{y}(\mathbf{r}_3)- \frac{1}{12} A'_{y}(\mathbf{r}_4)- \frac{1}{12} A'_{z}(\mathbf{r}_4)+ \frac{1}{12} A'_{y}(\mathbf{r}_5)+ \frac{1}{12} A'_{y}(\mathbf{r}_6)+ \frac{1}{12} A'_{z}(\mathbf{r}_6)$}\\ & & \multicolumn{8}{l}{$+ \frac{1}{12} A'_{y}(\mathbf{r}_7)+ \frac{1}{12} A'_{y}(\mathbf{r}_8)- \frac{1}{12} A'_{z}(\mathbf{r}_8)\left. \right ]$}\\
$A_{zz}$ & $ = $ & \multicolumn{8}{l}{$ \frac{1}{d_{\mathrm c}} \left [ \right. - \frac{1}{8} A'_{z}(\mathbf{r}_1)- \frac{1}{8} A'_{z}(\mathbf{r}_2)- \frac{1}{8} A'_{z}(\mathbf{r}_3)- \frac{1}{8} A'_{z}(\mathbf{r}_4)+ \frac{1}{8} A'_{z}(\mathbf{r}_5)+ \frac{1}{8} A'_{z}(\mathbf{r}_6)+ \frac{1}{8} A'_{z}(\mathbf{r}_7)+ \frac{1}{8} A'_{z}(\mathbf{r}_8)\left. \right ]$}\\
$A_{xxx}$ & $ = $ & \multicolumn{8}{l}{$ \frac{1}{d_{\mathrm c}^2} \left [ \right. + \frac{1}{4} A'_{x}(\mathbf{r}_1)+ \frac{1}{8} A'_{z}(\mathbf{r}_1)+ \frac{1}{4} A'_{x}(\mathbf{r}_3)- \frac{1}{8} A'_{z}(\mathbf{r}_3)+ \frac{1}{4} A'_{x}(\mathbf{r}_5)- \frac{1}{8} A'_{z}(\mathbf{r}_5)+ \frac{1}{4} A'_{x}(\mathbf{r}_7)+ \frac{1}{8} A'_{z}(\mathbf{r}_7)\left. \right ]$}\\
$A_{xxy}$ & $ = $ & \multicolumn{8}{l}{$ \frac{1}{d_{\mathrm c}^2} \left [ \right. + \frac{1}{4} A'_{y}(\mathbf{r}_1)+ \frac{1}{8} A'_{z}(\mathbf{r}_2)+ \frac{1}{4} A'_{y}(\mathbf{r}_3)- \frac{1}{8} A'_{z}(\mathbf{r}_4)+ \frac{1}{4} A'_{y}(\mathbf{r}_5)- \frac{1}{8} A'_{z}(\mathbf{r}_6)+ \frac{1}{4} A'_{y}(\mathbf{r}_7)+ \frac{1}{8} A'_{z}(\mathbf{r}_8)\left. \right ]$}\\
$A_{xxz}$ & $ = $ & \multicolumn{8}{l}{$ \frac{1}{d_{\mathrm c}^2} \left [ \right. - \frac{9}{80} A'_{x}(\mathbf{r}_1)+ \frac{1}{40} A'_{z}(\mathbf{r}_1)- \frac{1}{80} A'_{y}(\mathbf{r}_2)- \frac{1}{40} A'_{z}(\mathbf{r}_2)+ \frac{9}{80} A'_{x}(\mathbf{r}_3)+ \frac{1}{40} A'_{z}(\mathbf{r}_3)+ \frac{1}{80} A'_{y}(\mathbf{r}_4)- \frac{1}{40} A'_{z}(\mathbf{r}_4)+ \frac{9}{80} A'_{x}(\mathbf{r}_5)$}\\ & & \multicolumn{8}{l}{$+ \frac{1}{40} A'_{z}(\mathbf{r}_5)+ \frac{1}{80} A'_{y}(\mathbf{r}_6)- \frac{1}{40} A'_{z}(\mathbf{r}_6)- \frac{9}{80} A'_{x}(\mathbf{r}_7)+ \frac{1}{40} A'_{z}(\mathbf{r}_7)- \frac{1}{80} A'_{y}(\mathbf{r}_8)- \frac{1}{40} A'_{z}(\mathbf{r}_8)\left. \right ]$}\\
$A_{xyy}$ & $ = $ & \multicolumn{8}{l}{$ \frac{1}{d_{\mathrm c}^2} \left [ \right. + \frac{1}{8} A'_{z}(\mathbf{r}_1)+ \frac{1}{4} A'_{x}(\mathbf{r}_2)- \frac{1}{8} A'_{z}(\mathbf{r}_3)+ \frac{1}{4} A'_{x}(\mathbf{r}_4)- \frac{1}{8} A'_{z}(\mathbf{r}_5)+ \frac{1}{4} A'_{x}(\mathbf{r}_6)+ \frac{1}{8} A'_{z}(\mathbf{r}_7)+ \frac{1}{4} A'_{x}(\mathbf{r}_8)\left. \right ]$}\\
$A_{xyz}$ & $ = $ & \multicolumn{8}{l}{$ \frac{1}{d_{\mathrm c}^2} \left [ \right. - \frac{1}{16} A'_{y}(\mathbf{r}_1)- \frac{1}{16} A'_{x}(\mathbf{r}_2)+ \frac{1}{16} A'_{y}(\mathbf{r}_3)+ \frac{1}{16} A'_{x}(\mathbf{r}_4)+ \frac{1}{16} A'_{y}(\mathbf{r}_5)+ \frac{1}{16} A'_{x}(\mathbf{r}_6)- \frac{1}{16} A'_{y}(\mathbf{r}_7)- \frac{1}{16} A'_{x}(\mathbf{r}_8)\left. \right ]$}\\
$A_{xzz}$ & $ = $ & \multicolumn{8}{l}{$ \frac{1}{d_{\mathrm c}^2} \left [ \right. - \frac{1}{8} A'_{z}(\mathbf{r}_1)+ \frac{1}{8} A'_{z}(\mathbf{r}_3)+ \frac{1}{8} A'_{z}(\mathbf{r}_5)- \frac{1}{8} A'_{z}(\mathbf{r}_7)\left. \right ]$}\\
$A_{yyy}$ & $ = $ & \multicolumn{8}{l}{$ \frac{1}{d_{\mathrm c}^2} \left [ \right. + \frac{1}{4} A'_{y}(\mathbf{r}_2)+ \frac{1}{8} A'_{z}(\mathbf{r}_2)+ \frac{1}{4} A'_{y}(\mathbf{r}_4)- \frac{1}{8} A'_{z}(\mathbf{r}_4)+ \frac{1}{4} A'_{y}(\mathbf{r}_6)- \frac{1}{8} A'_{z}(\mathbf{r}_6)+ \frac{1}{4} A'_{y}(\mathbf{r}_8)+ \frac{1}{8} A'_{z}(\mathbf{r}_8)\left. \right ]$}\\
$A_{yyz}$ & $ = $ & \multicolumn{8}{l}{$ \frac{1}{d_{\mathrm c}^2} \left [ \right. - \frac{1}{80} A'_{x}(\mathbf{r}_1)- \frac{1}{40} A'_{z}(\mathbf{r}_1)- \frac{9}{80} A'_{y}(\mathbf{r}_2)+ \frac{1}{40} A'_{z}(\mathbf{r}_2)+ \frac{1}{80} A'_{x}(\mathbf{r}_3)- \frac{1}{40} A'_{z}(\mathbf{r}_3)+ \frac{9}{80} A'_{y}(\mathbf{r}_4)+ \frac{1}{40} A'_{z}(\mathbf{r}_4)+ \frac{1}{80} A'_{x}(\mathbf{r}_5)$}\\ & & \multicolumn{8}{l}{$- \frac{1}{40} A'_{z}(\mathbf{r}_5)+ \frac{9}{80} A'_{y}(\mathbf{r}_6)+ \frac{1}{40} A'_{z}(\mathbf{r}_6)- \frac{1}{80} A'_{x}(\mathbf{r}_7)- \frac{1}{40} A'_{z}(\mathbf{r}_7)- \frac{9}{80} A'_{y}(\mathbf{r}_8)+ \frac{1}{40} A'_{z}(\mathbf{r}_8)\left. \right ]$}\\
$A_{yzz}$ & $ = $ & \multicolumn{8}{l}{$ \frac{1}{d_{\mathrm c}^2} \left [ \right. - \frac{1}{8} A'_{z}(\mathbf{r}_2)+ \frac{1}{8} A'_{z}(\mathbf{r}_4)+ \frac{1}{8} A'_{z}(\mathbf{r}_6)- \frac{1}{8} A'_{z}(\mathbf{r}_8)\left. \right ]$}\\
    $A_{zzz}$ & $ = $ & \multicolumn{8}{l}{$ \frac{1}{d_{\mathrm c}^2} \left [ \right. + \frac{1}{16} A'_{x}(\mathbf{r}_1)+ \frac{1}{8} A'_{z}(\mathbf{r}_1)+ \frac{1}{16} A'_{y}(\mathbf{r}_2)+ \frac{1}{8} A'_{z}(\mathbf{r}_2)- \frac{1}{16} A'_{x}(\mathbf{r}_3)+ \frac{1}{8} A'_{z}(\mathbf{r}_3)- \frac{1}{16} A'_{y}(\mathbf{r}_4)+ \frac{1}{8} A'_{z}(\mathbf{r}_4)- \frac{1}{16} A'_{x}(\mathbf{r}_5)$}\\ & & \multicolumn{8}{l}{$+ \frac{1}{8} A'_{z}(\mathbf{r}_5)- \frac{1}{16} A'_{y}(\mathbf{r}_6)+ \frac{1}{8} A'_{z}(\mathbf{r}_6)+ \frac{1}{16} A'_{x}(\mathbf{r}_7)+ \frac{1}{8} A'_{z}(\mathbf{r}_7)+ \frac{1}{16} A'_{y}(\mathbf{r}_8)+ \frac{1}{8} A'_{z}(\mathbf{r}_8)\left. \right ]$}\\
    
    \hline
    $A_{x}(\mathbf{r}) $ & $ = $ & \multicolumn{8}{l}{$ A_{x}(\mathbf{r}_{\mathrm{cm}})  + A_{xxx}x^{\prime 2} + 2A_{xxy}x'y' + 2A_{xxz}x'z' + A_{xx}x' + A_{xyy}y^{\prime 2} + 2A_{xyz}y'z' + A_{xy}y' + A_{xzz}z^{\prime 2} + A_{xz}z' $}\\
    $A_{y}(\mathbf{r}) $ & $ = $ & \multicolumn{8}{l}{$ A_{y}(\mathbf{r}_{\mathrm{cm}})  + A_{xxy}x^{\prime 2} + 2A_{xyy}x'y' + 2A_{xyz}x'z' + A_{xy}x' + A_{yyy}y^{\prime 2} + 2A_{yyz}y'z' + A_{yy}y' + A_{yzz}z^{\prime 2} + A_{yz}z' $}\\
    $A_{z}(\mathbf{r}) $ & $ = $ & \multicolumn{8}{l}{$ A_{z}(\mathbf{r}_{\mathrm{cm}})  + A_{xxz}x^{\prime 2} + 2A_{xyz}x'y' + 2A_{xzz}x'z' + A_{xz}x' + A_{yyz}y^{\prime 2} + 2A_{yzz}y'z' + A_{yz}y' + A_{zzz}z^{\prime 2} + A_{zz}z' $}\\
    \hline
    $\mathbf{r}'_{\mathrm i}/d_{\mathrm c}$  & $=$ & $[\begin{matrix} 1 & 0 & -1 \end{matrix}]$  & $[\begin{matrix} 0 & 1 & -1 \end{matrix}]$  & $[\begin{matrix} -1 & 0 & -1 \end{matrix}]$  & $[\begin{matrix} 0 & -1 & -1 \end{matrix}]$  & $[\begin{matrix} 1 & 0 & 1 \end{matrix}]$  & $[\begin{matrix} 0 & 1 & 1 \end{matrix}]$  & $[\begin{matrix} -1 & 0 & 1 \end{matrix}]$  & $[\begin{matrix} 0 & -1 & 1 \end{matrix}]$  \\
    \hline\hline
  \end{tabular}
\end{table*}

Since all measure points can only obtain the long-range forces every \pkick~step, if the binary forms in the middle of \pdrift, it is not possible to construct the tidal-tensor field immediately.
Besides, if the binary disrupt, the tidal-tensor filed also cannot provide the correct perturbation once the two components leave far away.
This is the limitation of the tidal-tensor method.
However, the purpose of the tidal-tensor method is to ensure the long-term cumulative effect of long-range perturbation is correctly treated.
Thus, a short interval error within one $\dts$ is not very serious.

\subsubsection{Counter-force}

\paragraph{Orbit-sampling method}

The tidal-tensor method introduced above provide the correct perturbation to the internal motion of the binary, it is also necessary to ensure that the perturbers can obtain the consistent counter-force.
If the perturbers obtain the forces from the two components of the binary at \pkick, only one random phase of the binary is used to evaluate the interaction.
This cannot provide the correct orbit-averaged force from binaries.
To solve this issue, another group of artificial particles can be created by sampling the binary orbit with an equal eccentricity anomaly interval (Fig.~\ref{fig:tt}).
For example, if the eccentric anomaly interval, $\Delta\mathcal{E} = \pi/4$, $16$ particles are created along the two orbits of binary components.
The masses of these particles are weighted by the interval of mean anomaly:
\begin{equation}
  \begin{aligned}
  m_{\mathrm{orb},i,k} = & m_{i} \frac{\Delta \mathcal{M}_{k}}{2\pi}\\
  \Delta \mathcal{M}_{k} = & \Delta\mathcal{E} - e \left [ \sin \left(\mathcal{E}_{k} + \frac{\Delta\mathcal{E}}{2}  \right) - \sin \left( \mathcal{E}_{k} - \frac{\Delta\mathcal{E}}{2} \right) \right] \\
                         = & \Delta\mathcal{E} - 2 e \cos \left( \mathcal{E}_{k} \right) \sin \left( \frac{\Delta\mathcal{E}}{2} \right). \\
  \end{aligned}
  \label{orb:mw}
\end{equation}
where $\mathcal{M}_{k}$ and $\mathcal{E}_{k}$ are the mean anomaly and the eccentric anomaly at the point of particle $k$, respectively.
In this case, the particle mass approximately represents the orbit-average duration of the two-components at each $\Delta \mathcal{E}$ .

\paragraph{Pseudoparticle multipole method}

In the orbit-sampling method, at least $8$ sample particles are needed to reasonably represent the orbits of binaries.
It is rather expensive since the number of sample particles per binary is large.
\cite{Kawai2001} introduces the pseudoparticle multipole method that the quadrupole moment of a particle group can be represented by only three pseudoparticles.
The quadrupole tensor of $N$ particles can be described by
\begin{equation}
  \mathcal{A} = \sum_{i=1}^{N} {m_i \bm{r}_i \otimes \bm{r}_i},
  \label{eq:quad}
\end{equation}
where $\otimes$ is tensor production.
The corresponding traceless form iS
\begin{equation}
  \mathcal{A'} = \frac{3}{2} \mathcal{A} - \frac{1}{2} \tr (\mathcal{A}) 
               = \frac{3}{2} \sum_{i=1}^{N} {m_i \bm{r}_i \otimes \bm{r}_i} - \frac{1}{2} \sum_{i=1}^{N} {m_i \bm{r}_i^2},
  \label{eq:quadtr}
\end{equation}
where the term with $\bm{r}_i^2$ is subtracted from the the diagonal elements of the matrix by $\bm{r}_i \otimes \bm{r}_i$.

The orbit-average of the binary motion can be treated as a continue distribution of mass along the orbits of two components.
Thus we can also derive the analytic formulae of its quadrupole moment.
In the coordinate systems of the binary orbital plane where the three Delaunay's elements are zero, the relative position vector has the form depending on $\mathcal{E}$ as 
\begin{equation}
  \Delta \bm{r} = \left[\begin{matrix}a \left(\cos{\mathcal{E}} - e \right) & a \sqrt{1 - e^{2}} \sin{\mathcal{E}} & 0\end{matrix}\right].
  \label{eq:dr}
\end{equation}
The two component position vectors have a relation to $\Delta r$ by
\begin{equation}
  \begin{aligned}
    \bm{r}_1 &=  -\frac{m_2}{m_1+m_2} \Delta \bm{r}\\
    \bm{r}_2 &=   \frac{m_1}{m_1+m_2} \Delta \bm{r},\\
  \end{aligned}
  \label{eq:r1r2}
\end{equation}
where $m_1$ and $m_2$ are masses of two components.
Put Eq.~\ref{eq:dr} and~\ref{eq:r1r2} into Eq.~\ref{eq:quad}, we can obtain $\mathcal{A}$ of the binary as a function of $\mathcal{E}$.
When the two components pass one full orbit, $\mathcal{E}$ changes from $0$ to $2\pi$ and $t$ changes from $0$ to $P$ (period).
The orbital average of the quadrupole moment should integrate one period of $\mathcal{A}(\mathcal{E})$:
\begin{equation}
  \langle \mathcal{A} \rangle = \frac{1}{P} \int_0^{P} \mathcal{A}(\mathcal{E}) dt.
\end{equation}
The differentials of $t$ and $\mathcal{E}$ has the relation:
\begin{equation}
  \diff t = \frac{P}{2 \pi}(1 - e \cos{\mathcal{E}} ) \diff \mathcal{E}.
\end{equation}
Replace $\diff t$ by $\diff \mathcal{E}$ and do the integration, we can obtain the final form:
\begin{equation}
  \langle \mathcal{A} \rangle = \mu a^2 \left[\begin{matrix}2 e^{2} + \frac{1}{2} & 0 & 0\\0 & \frac{1}{2} - \frac{e^{2}}{2} & 0\\0 & 0 & 0\end{matrix}\right],
  \label{eq:avebinquad}
\end{equation}
where $\mu$ is reduced mass, $m_1m_2/(m_1+m_2)$.

We choose the coordinate system where $\langle \mathcal{A} \rangle$ becomes traceless in order to use the pseudoparticle multipole method:
\begin{equation}
  \begin{aligned}
    \langle  \mathcal{A'} \rangle =  &\frac{3}{2}\langle \mathcal{A} \rangle - \frac{1}{2} \tr(\langle \mathcal{A} \rangle) \\
                                  =  & \frac{1}{4} \mu a^2 \left[\begin{matrix}9 e^{2} + 1 & 0 & 0\\0 & 1 - 6 e^{2} & 0\\0 & 0 & - 3 e^{2} - 2\end{matrix}\right]\\
  \end{aligned}
\end{equation}

Using Eq.~3, 5 and 6 in \cite{Kawai2001}, the three pseudoparticles with the equal mass of $(m_1+m_2)/3$ are distributed at:
\begin{equation}
  \begin{aligned}
    \bm{r}_{\mathrm{p,1}} = & a\sqrt{\frac{\mu}{m_{1} + m_{2}}} \left[\begin{matrix}0 & \sqrt{1-e^{2}} & 0\end{matrix}\right]\\
    \bm{r}_{\mathrm{p,2}} = & a\sqrt{\frac{\mu}{m_{1} + m_{2}}} \left[\begin{matrix}\sqrt{3 e^{2} + \frac{3}{4}} & - \sqrt{\frac{1-e^{2}}{4}} & 0\end{matrix}\right]\\
    \bm{r}_{\mathrm{p,3}} = & a\sqrt{\frac{\mu}{m_{1} + m_{2}}} \left[\begin{matrix}-\sqrt{3 e^{2} + \frac{3}{4}} & - \sqrt{\frac{1-e^{2}}{4}} & 0\end{matrix}\right].\\
  \end{aligned}
  \label{eq:psep}
\end{equation}
These positions refer to the rest frame of the binary orbital plane.
To obtain the correct direction of the orbital plane in the original frame, we need to multiply Eq.~\ref{eq:psep} by the rotational matrix based on the Delaunay's elements.

\subsubsection{Test}
\label{sec:tttest}

To confirm that the artificial particle algorithm can provide the correct perturbation and counter-force,
we test a triple system with the initial condition listed in Table~\ref{tab:tt-test}.
\begin{table*}
  \caption{The initial condition of the hierarchical triple system for testing the artificial particle algorithm. $\mpr$ and $\ms$ are the masses of the primary and the secondary of the inner and outer binaries. The values are shown in the scale-free unit with the gravitational constant, $G=1$. $a$ is semi-major axis. $e$ is eccentricity. $\mathcal{I}$, $\phi$ and $\psi$ are Delaunay's elements. $\mathcal{E}$ is eccentric anomaly. $\tbin$ is period.}
  \begin{tabular}{ccccc ccccc}
    \hline
          & $\mpr$ & $\ms$   & $a$   & $e$    & $\mathcal{I}$ & $\phi$ & $\psi$ & $\mathcal{E}$ & $\tbin$ \\
    \hline
    in    & 0.00900& 0.00100 & 0.900 & 0.900  & 1.500 & 0.100 & 0.200 & 3.14 & $1.97\times10^{-3}$ \\
    out   & 1.00   & 0.01    & 1.500 & 0.0100 & 0.100 & 0.100 & 0.100 & 1.50 & 11.5 \\
    \hline
  \end{tabular}
  \label{tab:tt-test}
\end{table*}
We use three methods to integrate the motion of the system.
Orbits integrated by the accurate SDAR method is used as a reference (names as the SDAR-REF model).
Models using the \pppt~method with no artificial particles (no-TT), second- (TT-2) and third-order (TT-3) tidal-tensor methods are compared.
The perturber is outside the changeover region of the binary.
Thus the outer orbit is integrated by the leapfrog method.
The inner binary is integrated by the SDAR method.
$\dts=0.00390625$ for the \pppt~method.
The ratio between the binary period ($\tbinin$) and $\dts$ is about $0.51$.
Thus in the no-TT model, the long-range force is evaluated once every two $\tbinin$.
We also add a model of a binary with the initial condition the same as that of the outer binary in Table~\ref{tab:tt-test} (named as the B-out model).
The leapfrog integrator with the same step size is used.
The pseudoparticle multipole method is used for both TT-2 and TT-3 methods.

\begin{figure}
  \centering
  \includegraphics[width=1.0\columnwidth]{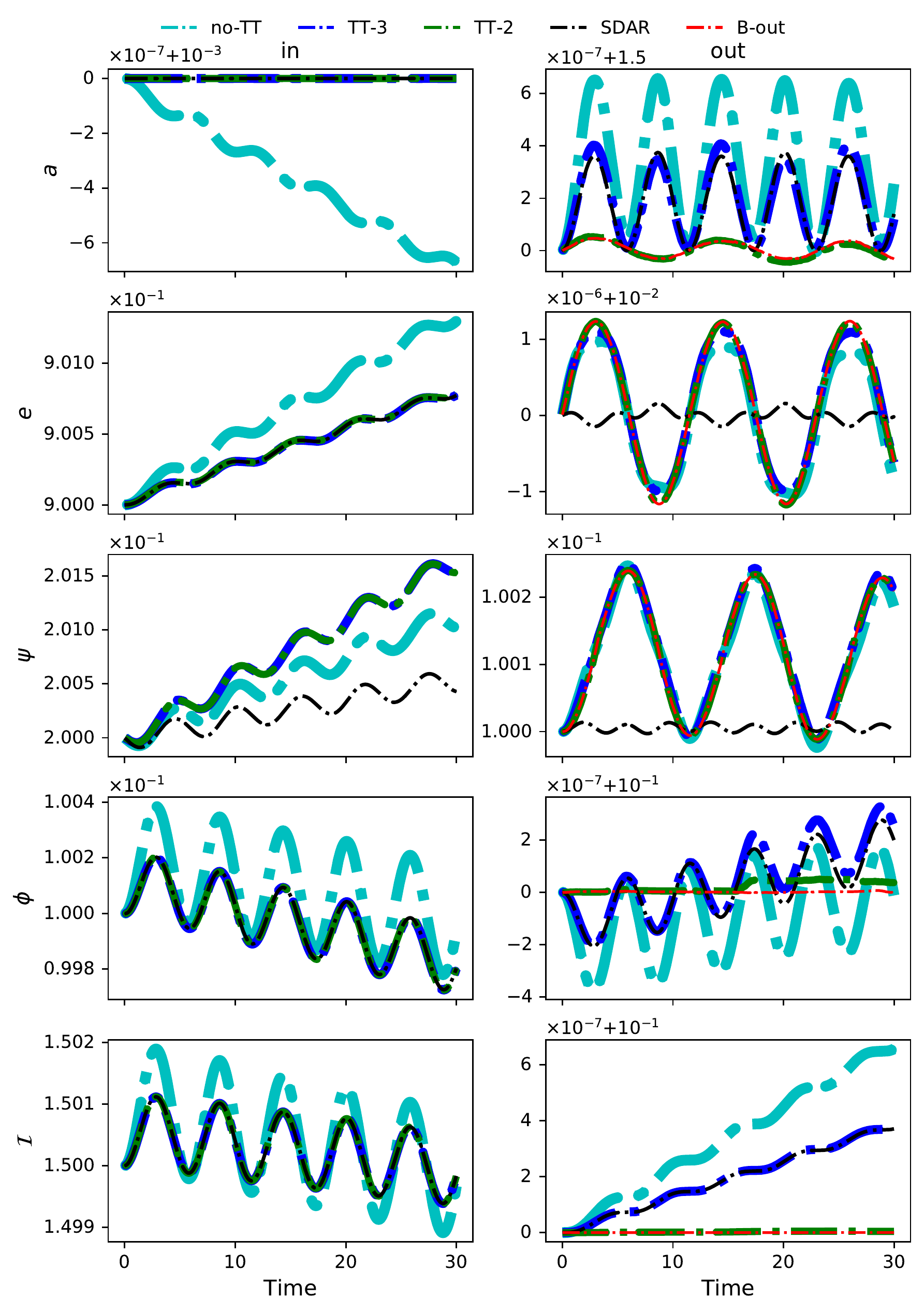}
  \caption{The evolution of orbital parameters of the inner and outer binaries for the triple system using different integration methods.
    For $a$ and time, the scale-free units are used.
    The black colour represents the accurate result using the SDAR method as a reference.
    The green and blue colours represent the TT and no-TT models, respectively.
    The purple colour represents the B-out model.
    For each panel, we apply the scientific notation in the plotting style of $y$-axis: the actual values of $y$-axis are calculated by $y_{\mathrm{tick}}\times \text{scale} + y_{\mathrm{offset}}$, where $y_{\mathrm{tick}}$ is the value shown along the the y-axis, $\text{scale}$ is the first value shown above the y-axis ($\text{scale}=1$ in default) and $y_{\mathrm{offset}}$ is the second value following the symbol ``+'' ($y_{\mathrm{offset}}=0$ in default). 
  }
  \label{fig:res-tt}
\end{figure}

The evolution of orbital elements are show in Fig.~\ref{fig:res-tt}.
By selecting the Cartesian coordinate system ($x$-$y$-$z$), the three Delaunay's elements (angles) are \citep[e.g.][]{Wang2020}:
\begin{itemize}
\item $\mathcal{I}$: inclination.
\item $\phi$: longitude of the ascending node.
\item $\psi$: argument of periapsis.
\end{itemize}
Except $\psi_{\mathrm{in}}$, both TT-2 and TT-3 models agree well with the SDAR-REF model on the evolution of inner orbital elements while the no-TT model does not.
This suggests that the tidal-tensor method indeed provides a better result for the secular motions of the inner binary.

On the other hand, the TT-3 model also provides a correct evolution of $a_{\mathrm{out}}$, $\phi_{\mathrm{out}}$ and $i_{\mathrm{out}}$.
But TT-2 model cannot reproduce the correct oscillation of the outer orbit and the evolution overlaps with the B-out model.
This suggests that the second-order tidal-tensor method cannot properly reproduce the secular motion of the outer orbit.
The no-TT model disagree with all others.

However, all three models show large differences (oscillation) of $e_{\mathrm{out}}$ and $\psi_{\mathrm{out}}$ compared with those of the SDAR-REF model, but agree well with those of the B-out models.
The B-out model is a simple binary motion, thus $e_{\mathrm{out}}$ and $\psi_{\mathrm{out}}$ should not evolve in reality.
This indicates that the artificial oscillation is caused by the inaccuracy of the leapfrog method for the outer orbit.

This result suggests that the third-order tidal-tensor algorithm is a good choice to represent a reasonable secular motions of both inner and outer binaries.
The low accuracy of the leapfrog method results in a relative error of $10^{-6}$ in the evolution of $e_{\mathrm{out}}$ but the averaged value can converge to the correct one.

\begin{figure}
  \centering
  \includegraphics[width=0.8\columnwidth]{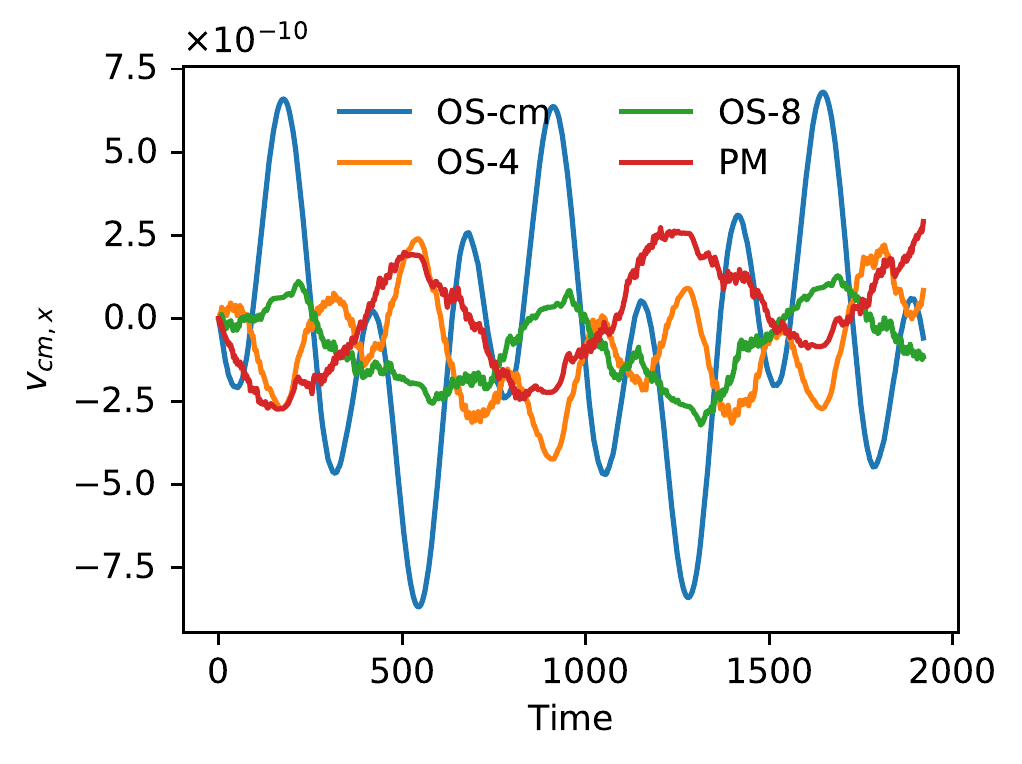}
  \caption{The evolution of the $x$-component of the centre-of-the-mass velocity of the triple system.
    The OS-cm model uses only the centre-of-the-mass of the inner binary to calculate the force to the perturber.
    The OS-4 and OS-8 models use orbit-sampling methods with $4$ and $8$ sample particles.
    The PM model uses three pseudoparticles.
  }
  \label{fig:vxorbit}
\end{figure}

How the counter forces are calculated does not affect the orbital motions as shown in Fig.~\ref{fig:res-tt}.
The results are identical within the resolution thus we does not show.
But the linear momentum conservation is sensitive to it.
In Fig.~\ref{fig:vxorbit}, the evolution of the $x$-component in the centre-of-the-mass velocity of the triple is shown.
Four models with different ways to calculate the counter forces are compared.
As the number of sample particles decrease, the error (oscillation) is more obvious.
The PM method provide a similar level of error as $8$ sample particles.
Thus it is a more efficient choice for the counter force if the high-order momentum is not important.

%

\subsection{Parallelization algorithm}

Fig.~\ref{fig:petar} shows how the hybrid parallelization is implemented.
The domain decomposition, exchanging particles between MPI processes, PT construction and long-range force calculations are handled by \fdps.
The MPI and OpenMP are used together in \fdps~with a well controlled load balance.
The clustering (Section~\ref{sec:clustering}) is also parallelized by using MPI and OpenMP methods \citep{Iwasawa2017}.
The long-range force calculation can be accelerated by both the SIMD instruction set of X86 architecture and the NVIDIA GPU using the CUDA programming environment \citep{Iwasawa2020}.
The AVX, AVX2 and AVX-512 instructions are used for the SIMD accelerated implementation.

\section{Benchmark}
\label{sec:test}

We carry out benchmarks to compare the long-term evolution of star clusters with and without binaries, and show the computing performances by using the \petar~and \nbpp~codes.
A scaling test is also performed on the Cray XC50 supercomputer.

\subsection{Comparison with \nbpp}

\subsubsection{Performance on GPU-based Desktop}
\label{sec:perfdesk}

We compare the performances of simulating star clusters with different numbers of particles ($N=1000$, 10000, 100000, 1000000) by using the \petar~and \nbpp~codes.
The name of models are listed in Table~\ref{tab:inpar}.
The initial mass function (IMF) of \cite{Kroupa2001} ranging from $0.08~M_\odot$ to $40~M_\odot$ is applied.
The Plummer model is used to generate the positions and velocities.
The system is in virial equilibrium and the virial radius is $1.0$ in the \cite{Henon1971} unit (hereafter named as NB unit).
No tidal field and no stellar evolution are used in order to have a well-controlled comparison of the dynamics.
For each $N$, we have two models with and without primordial binaries.
In models with binaries, we use the period and eccentrmaricity distributions from \cite{Kroupa1995a,Kroupa1995b}.
Initially all particles are in binaries.
The period distribution of this model has the form:
\begin{equation}
  \mathcal{F}(\tbin) = 2.5 \frac{\log_{10}{\tbin} - 1}{45 + (\log_{10}{\tbin} - 1)^2},
\end{equation}
where the maximum and minimum of $\log_{10}{\tbin}$ are truncated at $8.43$ and $1.00$ (in the unit of days), respectively.
The eccentricity follows the thermal distribution.
The orbits of some tight binaries are adjusted to avoid stellar collision \citep[pre-main sequence eigenevolution;][]{Kroupa1995b}.
Thus, a wide range of $\tbin$ and eccentricities are covered, which is very suitable for testing the code.
We assume that the initial half-mass radii of all models are $1.0$~pc and the periods are scaled to be in the NB time unit.

The simulations are performed on a GPU-based desktop computer.
The computer is equipped with one AMD RYZEN 3970X CPU ($3.7$~GHz) which includes $32$ physical cores, one NVIDIA RTX 2080Ti GPU and 4-channel DDR4-3200 SDRAM memories.
Both the codes use the hybrid parallelization methods containing MPI, OpenMP, AVX2 and GPU (CUDA).
For small numbers of particles, using all CPU cores causes the issue of load balance and overshooting of communication.
Thus, the number of cores for each simulations are adjusted to obtain the best performance.

One important point is that the performance of the two codes are sensitive to different inputting parameters.
We adjust the parameters for each model in order to optimize the performance.
The important ones are listed in Table~\ref{tab:inpar}.

\begin{table*}
  \caption{The optimized sets of input parameters that influence the performance of \nbpp~and \petar~codes. $N$ is the total number of particles; $N_{\mathrm b}$ is the total number of binaries; $R_{\mathrm{KS}}$ and $\Delta t_{\mathrm{KS}}$ are the criterion to switch on the KS regularization in \nbpp; $\dts$ and $\routr$ are the tree time step and the outer boundary reference of the changeover function for \petar. NB units are used for distance and time.}
  \label{tab:inpar}
  \begin{tabular}{ccccccc}
    \hline 
    Model & $N [10^3]$ & $N_{\mathrm b} [10^3]$ & $R_{\mathrm{KS}} [10^{-3}]$ & $\Delta t_{\mathrm{KS}} [10^{-5}]$ & $\dts$ & $\routr [10^{-2}]$ \\
    \hline
    N1k   &1           & 0                    & 1                        & 3.2                             & $1/128$ & 3.22490  \\
    N10k  &10          & 0                    & 0.464                    & 1                               & $1/256$ & 1.61101  \\
    N100k &100         & 0                    & 0.215                    & 0.3                             & $1/512$ & 0.801104 \\
    N1m   &1           & 0                    & 0.1                      & 0.1                             & $1/1024$ & 0.399470 \\
    N1kb  &1           & 0.5                  & 1                        & 3.2                             & $1/256$ & 1.62651 \\
    N10kb &10          & 5                    & 0.764                    & 1                               & $1/512$ & 0.802076 \\
    N100kb&100         & 50                   & 0.515                    & 0.3                             & $1/1024$ & 0.400471 \\
    N1mkb &1000        & 500                  & 0.4                      & 0.1                             & $1/2048$ & 0.199039 \\
    \hline
    \multicolumn{2}{c}{The shared parameters} &  \multicolumn{2}{l}{ \nbpp} & \multicolumn{2}{l}{$\eta_{\mathrm r} = \eta_{\mathrm i} = 0.1\sqrt{2}$; $\Nbp=50$} \\
    \multicolumn{2}{c}{}                      &  \multicolumn{2}{l}{ \petar}& \multicolumn{2}{l}{$\eta = 0.1$; $\theta=0.3,0.5$; $\dfrac{\routr}{\rinr} = 10.0$; $3^{\mathrm{rd}}$ tidal-tensor; PM}\\
    \hline
  \end{tabular} 
\end{table*}

For the Hermite integrator of both two codes, the time step is calculated by \citep{Aarseth2003,Oshino2011}
\begin{equation}
    \dthi = \mathrm{min}\left ( \eta \sqrt{\tfrac{\sqrt{\left|\bm{A}^{(0)}_i\right|^2 + A^2_{0}} \left|\bm{A}^{(2)}_i\right| + \left|\bm{A}^{(1)}_i\right|^2}{\left|\bm{A}^{(0)}_i\right| \left|\bm{A}^{(3)}_i\right| + \left|\bm{A}^{(2)}_i\right|^2} }, \dthm \right),\\
  \label{eq:dth}
\end{equation}
where $\bm{A}^{(j)}_i$ is the acceleration of a particle ($i$) and its $j$-order time derivatives, $A_{0}$ is the constant coefficient for safety and $\dthm$ is input parameter.
In \nbpp, $A_{0}$ is calculated only for particles having no neighbours by assuming an artificial particle locating at the centre of the system with the mass of $0.01 \mave$.
In \petar, $A_{0} = 0.1 \mave / \rnbi^2$ where $\mave$ is the local averaged mass of particles.
Since both codes use the block-time-step method, the calculated $\dthi$ is adjusted to the integer power of $0.5$.
In \nbpp, $\eta_{\mathrm i}$ for neighbours and $\eta_{\mathrm r}$ for distance particles are set to $0.1\sqrt{2}$.

Another three major parameters in \nbpp~that determine the performance are the expected number of neighbours, $\Nbp$, the separation of a neighbour pair, $R_{\mathrm{KS}}$, and the Hermite time step to trigger on KS regularization, $\Delta t_{\mathrm{KS}}$.
In our models, $\Nbp=50$.
One important tip is that the KS criterion not only influences the performance, but also has a big impact on the accuracy of integrating the internal motions of binaries.
A strict criterion can avoid a too frequent switch of KS but increase the error of integrating orbits (especially for eccentricities) of wide binaries.
The choices of $R_{\mathrm{KS}}$ and $\Delta t_{\mathrm{KS}}$ are shown in Table~\ref{tab:inpar}.

In \petar, the Hermite time-step coefficient does not significantly influences the performance.
We use the value of $0.1$.
The major impact to the performance comes from $\theta$, $\dts$ and the changeover function.
We compare $\theta$ of $0.3$ and $0.5$ with quadrupole moment of the particle-tree force for all models.
The energy error of the \pppt~method depends on the combination of $\dts$ and the changeover function \citep{Iwasawa2015}.
When $\dts$ is given, we determine the reference of the outer changeover boundary as:
\begin{equation}
  \routr = 10 \dts \sigma_{\mathrm{1D}}
  \label{eq:routdts}
\end{equation}
where $\sigma_{\mathrm{1D}}$ is the one-dimensional velocity dispersion of the system.
We fix $\routr/\rinr$ to $10.0$.
The third-order tidal-tensor and pseudoparticle multipole methods are used.

\begin{figure}
  \centering
  \includegraphics[width=1.0\columnwidth]{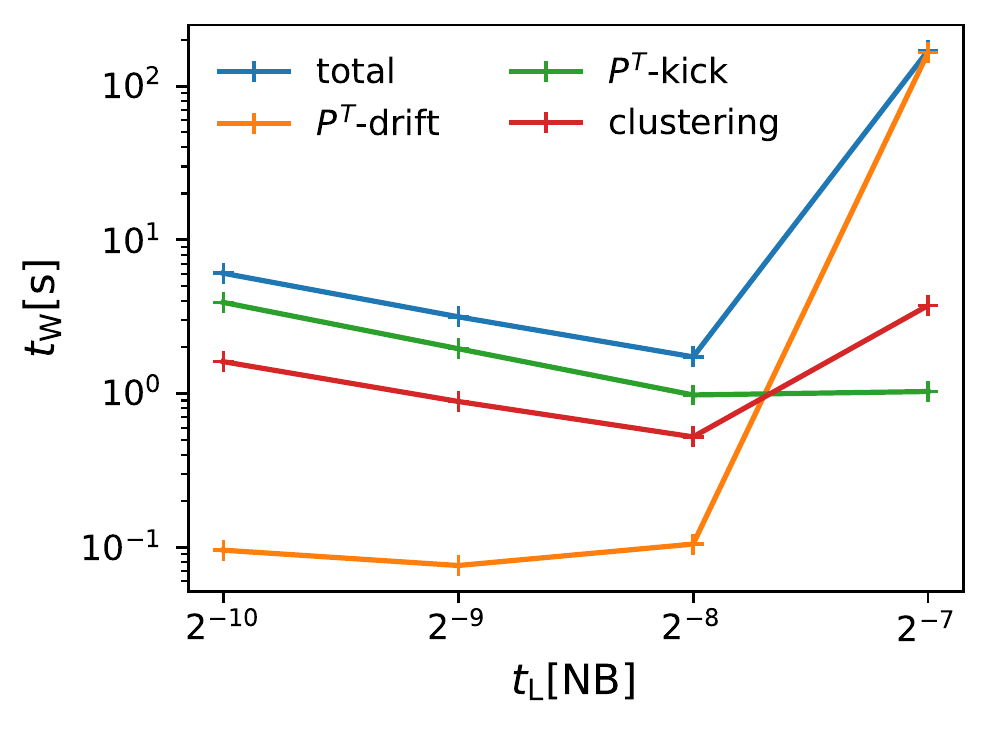}
  \caption{The wall clock times per NB time unit ($\tcomp$) depending on $\dts$ for the N10k model performed by using \petar.
    Colours represent the different parts of the computing (total, \pkick, \pdrift~and clustering).
    The minimum value of the total $\tcomp$ indicates the balanced choice of $\dts$ for the best performance.}
  \label{fig:finddt}
\end{figure}

By checking a group of $\dts$, we can find a balanced combination of the parameters to obtain the best performance.
The computational cost of the long-range force and kick velocity per \pkick~step is roughly constant.
Thus the wall clock time of one NB time unit ($\tcomp$) is anti-correlated with $\dts$.
In contrast, when $\dts$ is reduced, the sizes of clusters for short-range interactions are smaller due to a shorter $\routr$.
This gives a better load-balance and a less computational cost in the \pdrift~and the clustering.
Therefore, one balance $\dts$ can be found to achieve the best performance.
In \petar, we first estimate $\routr$ by
\begin{equation}
  \routr = 0.1 \frac{G M}{3 N^{1/3} \sigma_{\mathrm{1D}}^2},
\end{equation}
where $M$ is the total mass of the system and $G$ is gravitational constant.
Then using Eq.~\ref{eq:routdts}, we obtain the first guess of $\dts$ and check the best value around it.
One example of this check process for the N10k model is shown in Fig.~\ref{fig:finddt}.
When $\dts$ increases, $\tcomp$(\pkick) and $\tcomp$(clustering) decrease while $\tcomp$(\pdrift) increases.
The balanced $\dts=1/256$.
Above this value, $\tcomp$(\pdrift) increases significantly because a very large cluster with $2000$ members forms due to a set of large neighbour radii.
This completely kills the load-balance of paralellization and the benefit of the \pppt~method.

\begin{figure}
  \centering
  \includegraphics[width=1.0\columnwidth]{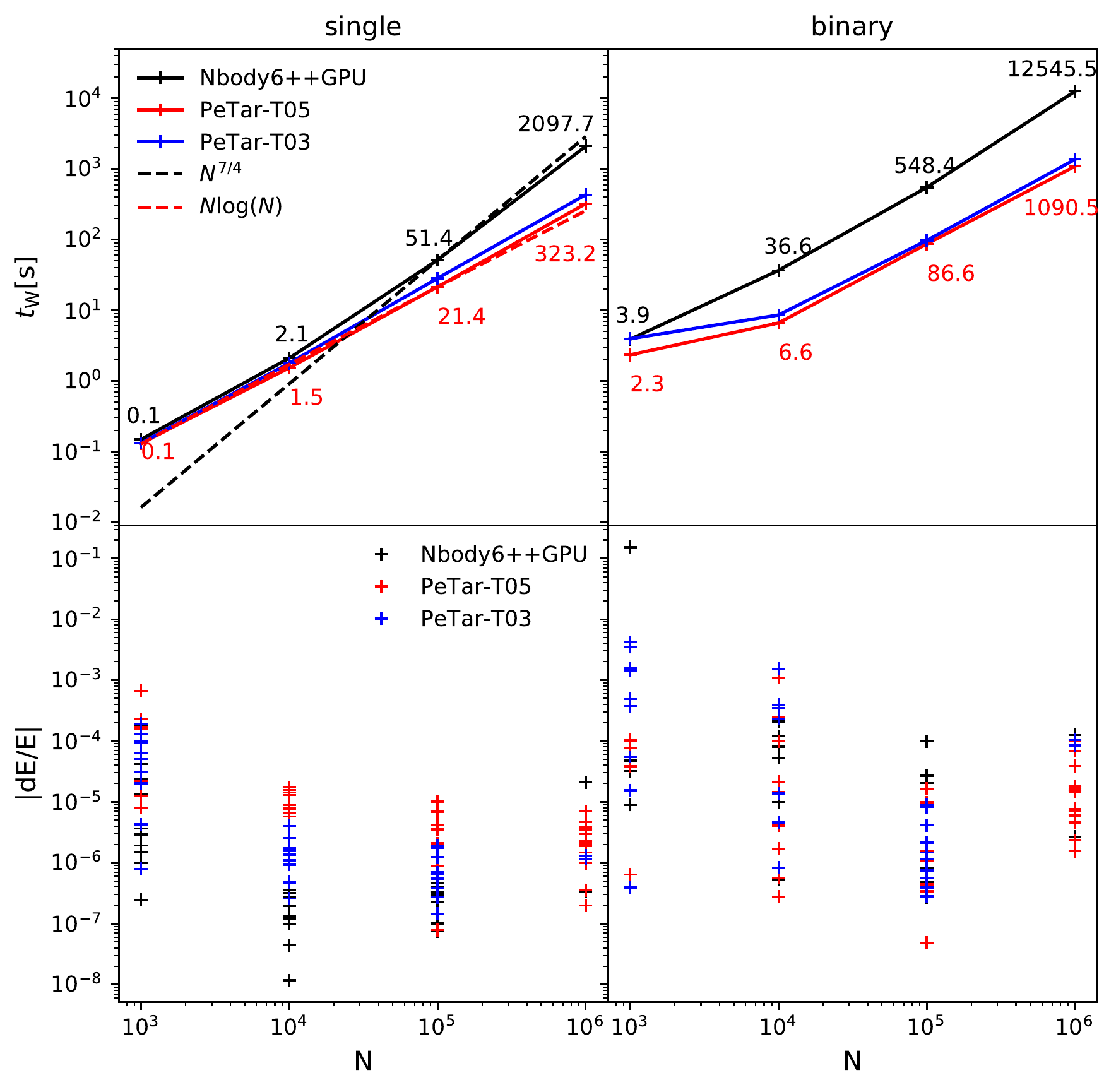}
  \caption{The comparison of wall clock time, $\tcomp$ (upper panel), and relative energy error, $\dEoE$ (lower panel), for models listed in Table~\ref{tab:inpar}.
    For the data of \petar, ``-T03'' and ``-T05'' indicate the opening angles of $0.3$ and $0.5$, respectively.
    The values of $\tcomp$ in sec are also printed near the data points.
    $\tcomp$ takes the average value of a few steps to avoid fluctuation.
    $\dEoE$ of each step (maximum $10$ steps) is shown.
    The left and right panels show the models without and with binaries, respectively.
  }
  \label{fig:perfdesk}
\end{figure}

The results of $\tcomp$ and the relative energy error ($\dEoE$) per NB time unit are shown in Fig.~\ref{fig:perfdesk}.
For each model, the maximum $10$ NB time steps are simulated to reduce fluctuation.
In both codes, the models with binaries sometimes have a very large $\tcomp$ in a short time interval due to the events of few-body interaction and the existence of semi-stable (perturbed) few-body systems.
In our analysis, we remove these data since they do not represent the normal performance of the codes.
Besides, we remove the first one steps in some models performed by \nbpp~if their $\tcomp$ are very different from that of other steps.
In the long-term simulations, the averaged $\tcomp$ may be larger if hierarchical systems frequently form.
This is easier to happen when $N$ is small, because the boundary of wide binaries (Eq.~\ref{eq:ab}) is large thus more space is available to form hierarchical systems.

For models without binaries, the two codes have a similar performance for $N$ of $1000$ and $10000$.
The differences appears when $N$ becomes large.
$\tcomp$ of \petar~well follows the scaling line of $N\log{N}$.
However, the result of \nbpp~scale differently for small and large $N$.
It is expected that $\tcomp$ is proportional to $N^{7/4}$, which is estimated by including the AC neighbour scheme.
Only when $N$ is large, the result follows $N^{7/4}$, while when $N$ is small, it follows $N\log{N}$.
The reason is probably due to the scaling of parallelization.
When $N$ is too small, the parellelization of multi cores does not help to improve the performance.
As $N$ increases, the computational cost of direct $N$-body method significantly increase, thus the parallelization efficiency (floating point operations vs. peak performance of CPU and GPU) also increases.
Once the efficiency reaches the maximum, the scaling begins to follow $N^{7/4}$.
In million-body case, \petar~provides a five times faster performance than that of \nbpp.

The significantly difference of the performance is shown in the models with large number of primordial binaries ($100\%$).
For $N$ of $1000$, the two codes have a similar performance, while the difference starts from $N=10000$.
\petar~code gives a much faster performance compared to \nbpp.
In million-body case, the performance difference is $12$ times.
As discussed in Section~\ref{sec:bottleneck}, \nbpp~does not parallize the KS regularzation and the switching of KS is very expensive.
This significantly reduces the performance.
In contrast, the behaviour of \petar~is much better.
Especially, the actual computing time of million-body model with full of binaries by using \petar~is even faster than the million-body model without binaries performed by \nbpp.
This result show the great advantage of \petar~for large $N$-body models with many binaries.

The relative energy errors, $\diff E/E$, are also compared in the bottom panels of Fig.~\ref{fig:perfdesk}.
The fluctuation is large for both codes.
In the models with no binaries, \nbpp~gives a systematically better energy conservation for $N<10^5$.
The small $\theta$ help to reduce the errors while the performance does not significantly change.
The wide range of mass spectrum is one important reason for the larger error.
With equal masses, the error is much smaller (not shown here).
The major contribution of the error in the models performed by \petar~comes from the \pkick, due to the approximation of the long-range force.

For a particle group, the changeover radii of Eq.~\ref{eq:rcut} takes care of the mass-dependent long-range tidal perturbation.
However, the mass-dependent error still exists in the centre-of-the-mass motion of the group due to the low accuracy of the leapfrog method (see Section~\ref{sec:tttest}).
Especially, the gravitational focusing enhances the error of the massive objects if the changeover radii is not sufficient large.
To avoid this, we need a $(m_i/\mave)^{1/2}$ dependent changeover radii, which can cause the formation of large clusters and the load-balance can become bad.
One possible solution is to implement the nested OpenMP parallelization for the Hermite and SDAR integrations inside the clusters.
This cannot fully solve the load-balance issue but may provide a better performance than reducing $\dts$ or $\theta$ in some conditions, especially when a very massive object like SMBH exists.

Nevertheless, both the codes show an increasing cumulative errors of a comparable level in the long-term simulation, while the dynamical behaviour seems to be consistent (Section~\ref{sec:longterm}).
If the global trend is not bad, even the energy error suddenly increases in one step, it only indicates that one specific event (in most cases this is a few-body interaction) is not well treated, but the global evolution is still statistically correct.

In the models with binaries, both codes have a similar level of errors.
However, we should notice that the definition of error with binaries are different in \nbpp~and \petar.
In \nbpp~the total energy of the systems are included in the energy conservation check, thus the energy error is completely dominated by the binary with the highest binding energy.
In \petar, we follow the definition of slow-down energy used in the \sdar code \citep{Wang2020}, where the energy of binaries are scaled down by the slow-down factor.
Thus the energy error reflects more about the global behaviour of the system.

\subsubsection{Long-term evolution}
\label{sec:longterm}

\paragraph{Lagrangian and core radii}

We check the long-term evolution including the post-core collapse stage of a star cluster without binaries (N250k model).
Initially, the cluster contains $250,000$ stars with the initial mass function (IMF) of \cite{Kroupa2001} ranging from $0.08~M_\odot$ to $40~M_\odot$.
For \nbpp, the initial $\Nbp=200$, $R_{\mathrm{KS}}=1.0\times 10^{-5}$ and $\Delta t_{\mathrm{KS}}=1.0\times 10^{-7}$ while the values are adjusted in the middle of the simulation.
For \petar, $\dts=1/1024$, $\routr = 0.00398365$ and $\theta=0.5$.

\begin{figure}
  \centering
  \includegraphics[width=1.0\columnwidth]{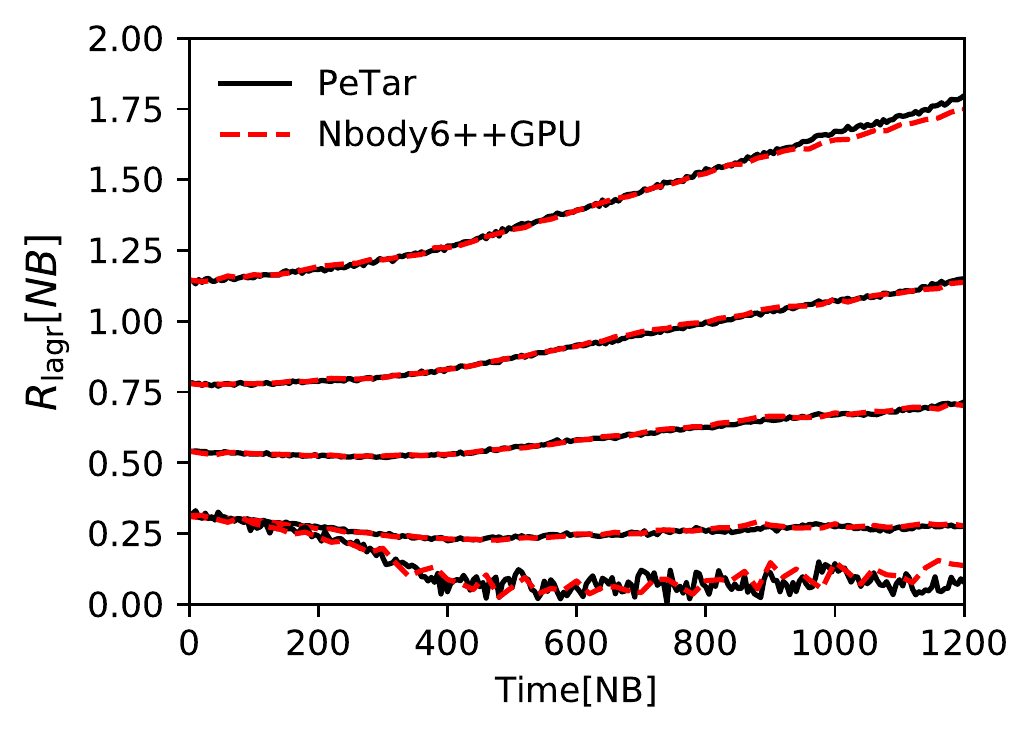}
  \caption{The evolution of the Lagrangian and core radii (N250k model) by using \petar~(black solid curves) and \nbpp~(red dashed curves). From the bottom to the top, the curves represent the core radius, $10\%$, $30\%$, $50\%$ $70\%$ Lagrangian radii, respectively.}
  \label{fig:lagr}
\end{figure}

The evolution of the Lagrangian and core radii of the N250k model is shown in Fig.~\ref{fig:lagr}.
We can see that the results of the two codes well overlap each others.
The core collapse finishes around $300$ NB time unit.
Both codes give the same core-collapse time and the evolution of the core radius.
This result indicates that \petar~can provide the same long-term evolution of the global density profile as the direct $N$-body method.

\paragraph{Escapers} 

\begin{figure}
  \includegraphics[width=1.0\columnwidth]{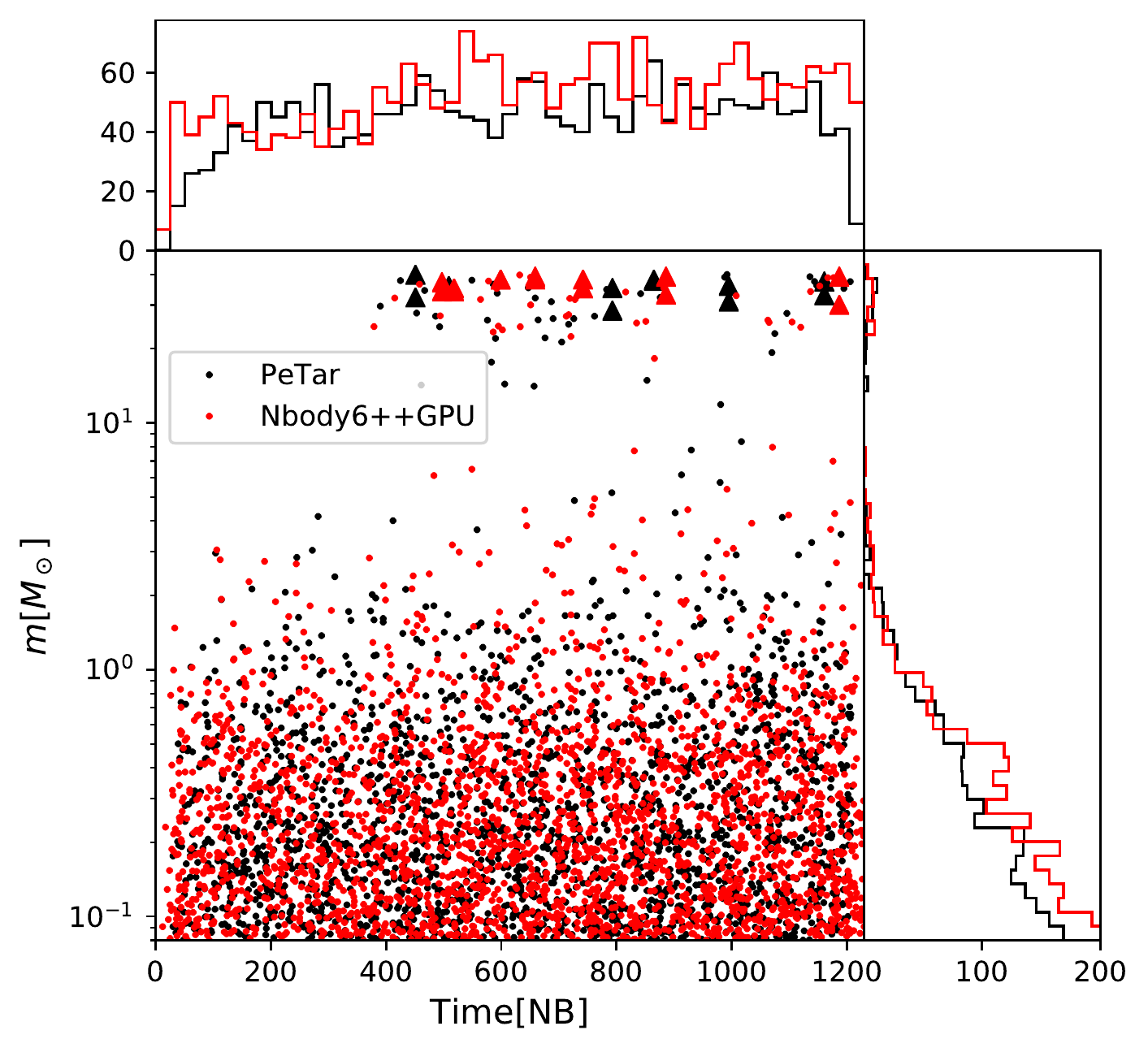}\\
  \includegraphics[width=0.8\columnwidth]{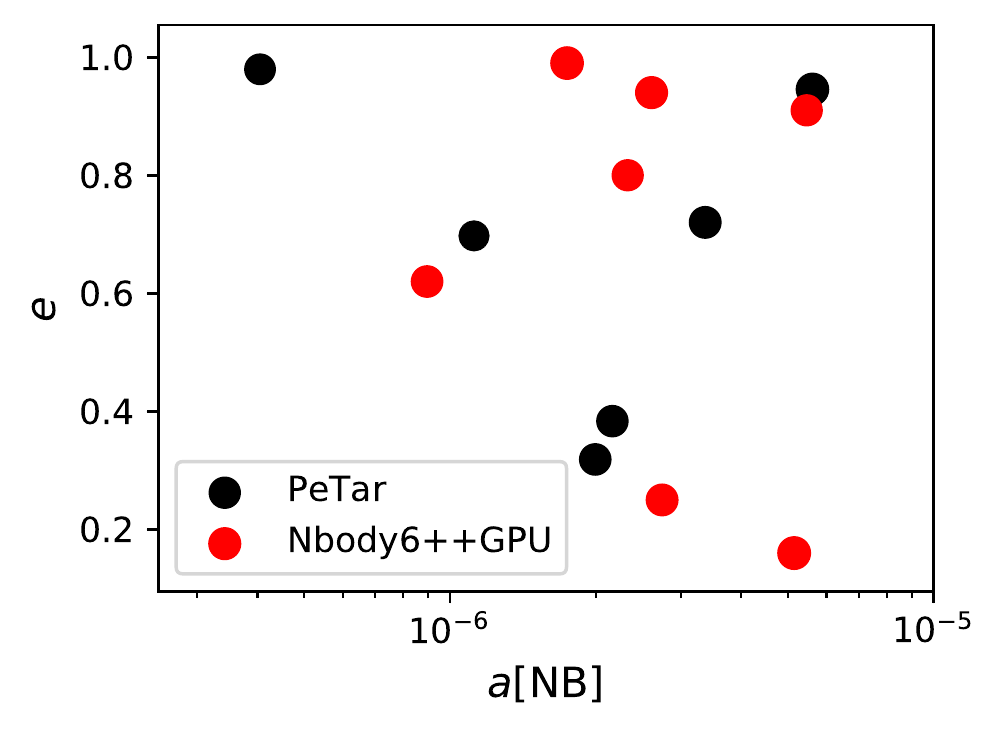}\\
  \caption{The properties of escapers for the N250k model performed by using \petar~and \nbpp. The upper panel: time and mass distribution of single and binary escapers. Triangles indicate the two components of each binary escaper.
    The mass is scaled to the unit of solar mass.
    The lower panel: semi-major axis ($a$) and eccentricity ($e$) of binary escapers. The areas of markers are proportional to the masses of binaries}
  \label{fig:N250esc}
\end{figure}

We compare the properties of escapers for the N250k model in Fig.~\ref{fig:N250esc}.
Two codes provide consistent time and mass distribution of escapers.
We can also clearly identify the low-mass escapers caused by the relaxation driven evaporation and the high-mass escapers ejected by strong few-body interactions in the core.
The latter appears after the core collapse.
The two components of binary escapers have similar masses and are the most massive objects in the cluster.
Both codes provide a similar number of binary escapers ($6$ and $7$) with very similar masses and consistent distribution of $a$ and $e$.
This result indicates that \petar~and \nbpp~well agree on the properties of escapers for the model without primordial binaries.

\paragraph{Performance and energy error} 

\begin{figure}
  \centering
  \includegraphics[width=1.0\columnwidth]{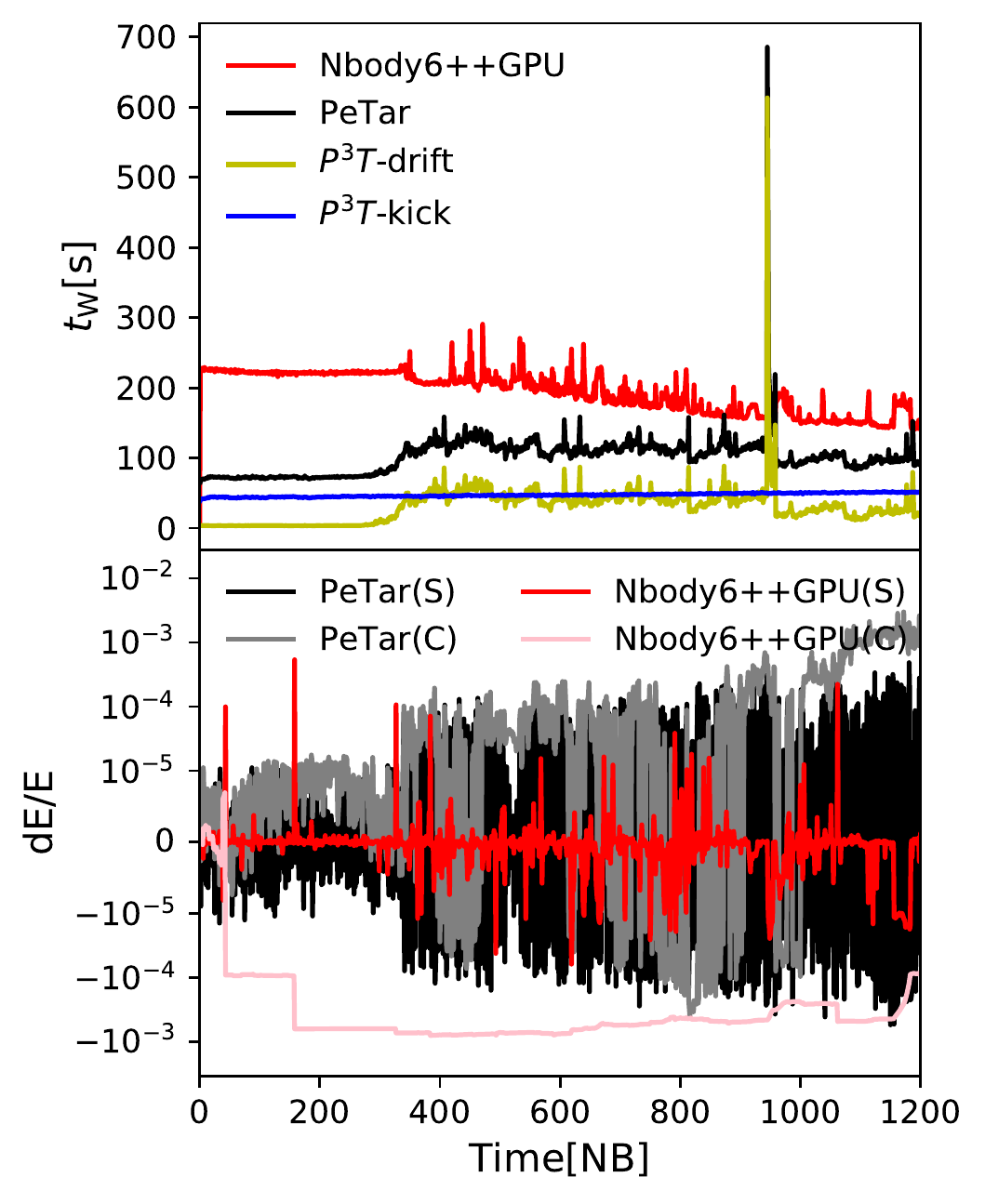}
  \caption{$\tcomp$ of total, \pkick~and \pdrift~(upper panel) and $\diff E/E$ (lower panel) vs. the physical evolution time for the N250k model. The label, (S) and (C), in the legend of $\diff E/E$ indicate ``per time'' and ``cumulative'', respectively.
  }
  \label{fig:perflong}
\end{figure}

In Fig.~\ref{fig:perflong}, the long-term behaviour of the performance and the energy error of the N250k model are compared.
At the beginning, \petar~provides a twice faster performance than that of \nbpp.
After core collapse, the performance of \petar~slows down due to a larger cost of \pdrift~(yellow curves).
This is expected since a dense core results in a larger size of  particle cluster and smaller Hermite time steps.
There are a few sharp peaks, which are due to an expensive calculation of a few-body interaction.
Such behaviour can happen when a tight multiple system (e.g. triple, quadruple) forms and stay for a while before a disruption by close encounters.
Sometimes, the criterion for switching on and off the SDAR method or the initial integration step size are not well adjusted to the specific condition.
Thus a larger error or an expensive calculation can appear.
There is probably no uniform way that can well handle all type of few-body interactions.
This is the same for both \nbpp~and \petar.
The performance of \nbpp~becomes better in the late phase of the evolution.

The relative energy error $\diff E/E$ per step (S) of \petar~has a larger fluctuation while the cumulative errors (C) converge around zero until $t=1000$~NB unit.
Large fluctuations appear after the core collapse when few-body interactions start to eject massive objects out of the core.
The appearance of large $\diff E/E$ at one step is often associated with the formation of a large particle cluster including a few hundreds members.
This is caused by a high-velocity particle which has a large $\rnbi$.
In such case, the particle moves a long distance during one $\dts$, thus the time step of the long-range interaction is too large.
If this particle is also massive, a large $\diff E/E$ can appear.
The way to solve this issue is to reduce $\dts$.
In our simulation, to have consistent parameters for measuring the performance, $\dts$ is not modified.
In an application of an astrophysical study, after core collapse, it is better to adjust the $\dts$ and $\routr$ based on the new state of the system.

The behaviour of $\diff E/E$ by using \nbpp~is better but large jumps of $\diff E/E$ sometimes appear and result in a large cumulative error.
From the information record of the simulations, significant time-step jumps happen for some particles when the big error appears.
It can be caused by the unsuitable criterion to switch on KS regularization, that a strong close encounter is not caught by the KS method.
However, it is not easy to design a general criterion to handle all kind of case while keeping a good performance.
In such case, the usual way to avoid such large error is to restart the simulation by reducing $\eta_{\mathrm r}$ and $\eta_{\mathrm i}$ or enlarging the KS criterion, $R_{\mathrm{KS}}$ and $\Delta t_{\mathrm{KS}}$.
For a test purpose, we do not restart the simulation and allow such large $\diff E/E$(S) in our models.

Although the error of $10^{-3}$ is not small, the behaviour of the long-term dynamical evolution is not very sensitive to it, as shown in Fig.~\ref{fig:lagr}.
This suggests that we should not use the energy conservation as a strict judgement for the quality of simulations, but should focus on what the physical processes we really care and whether they are treated properly.
If the interested objects cause a big energy error, we need to properly validate the result with better energy conservation.
On the other hand, sometimes energy conservation can be misleading when we compare the simulations done by using different approximations.
For instance, if all binaries in the clusters are treated as isolated and do not exchange energy with other stars, the energy conservation can be much better since the difficulty of the few-body interactions is avoided, but the results are completely wrong.
Another example is to use the softening length, which gives a wrong behaviour of close encounters but result in a much better energy conservation.

\subsubsection{Binaries}

\begin{figure}
  \centering
  \includegraphics[width=1.0\columnwidth]{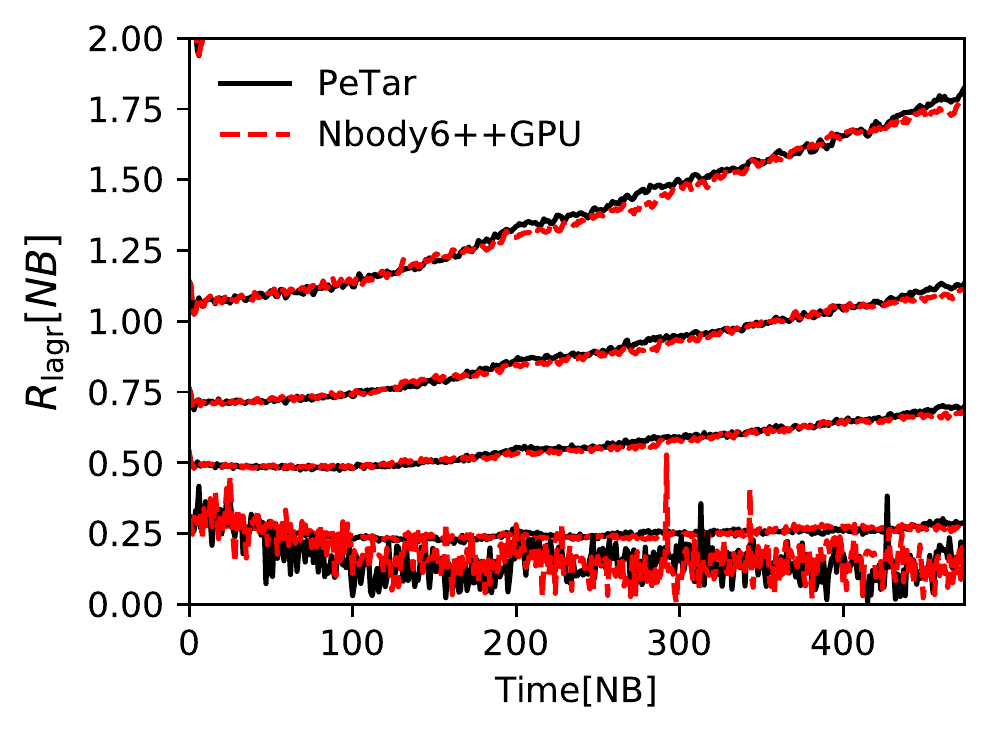}
  \caption{The evolution of the Lagrangian and core radii for N100kb model. The plotting style is similar to Fig.~\ref{fig:lagr}.}
  \label{fig:lagrb}
\end{figure}

\begin{figure}
  \centering
  \includegraphics[width=1.0\columnwidth]{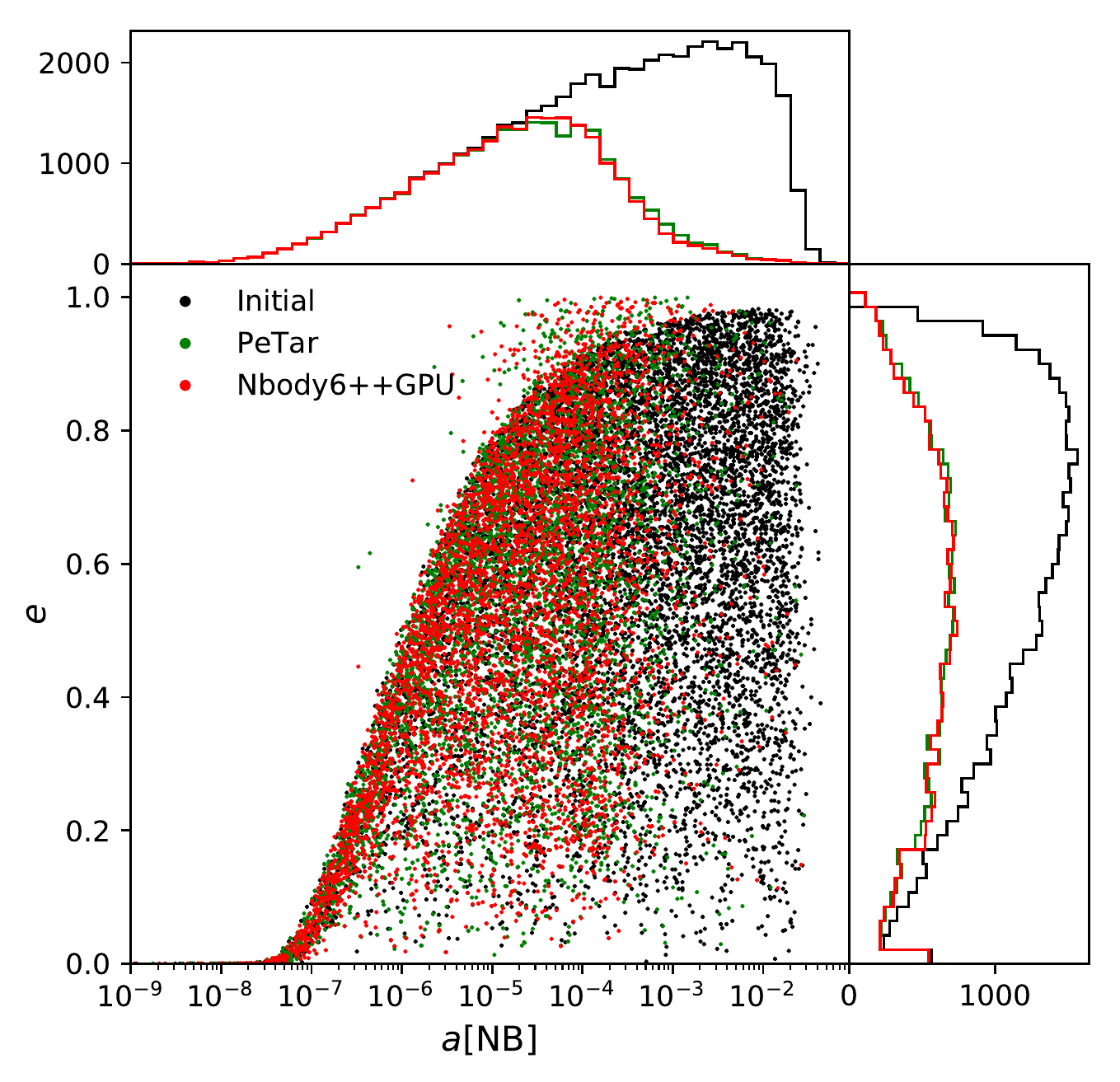}
  \caption{Distributions of the semi-major axis $a$ and eccentricity $e$ for the N100kb model performed by using \petar~and \nbpp. Each point in the central panel indicate one binary. We randomly select $20\%$ binaries to show in order to avoid too many data points. The top and right panels show the histograms of $a$ and $e$, respectively. The black, green and red colours indicate the initial condition, the results of \petar~and \nbpp, respectively.}
  \label{fig:binN100kb}
\end{figure}

We perform another simulation of star clusters with primordial binaries (the N100kb model) in order to compare the behaviour of the dynamical evolution of binaries.
The initial controlling parameters are the same as shown in Table~\ref{tab:inpar}.
For \petar, we choose the opening angle as $\theta = 0.3$ for a better accuracy.
The evolution of Lagrangian and core radii are shown in Fig.~\ref{fig:lagrb}.
Similar to the result of N250k, the two results agree with each other very well.

The initial and final distribution (after $475$ NB time unit) of semi-major axis, $a$, and eccentricity, $e$, are compared in Fig.~\ref{fig:binN100kb}.
There is a forbidden region of $a$ and $e$ in the initial distribution based on the pre-main-sequence eigenevolution model of \citep{Kroupa1995b}.
Since stars have physical sizes, the pericentres of binaries cannot be lower than a threshold, otherwise the binaries collide.
In our simulations, particles are treated as point masses.
Thus, after interactions, a part of binaries, especially the tight binaries which are easily perturbed, can enter the forbidden region.
The two histograms show that the two codes provide very similar final distributions of $a$ and $e$.

\begin{figure}
  \centering
  \includegraphics[width=1.0\columnwidth]{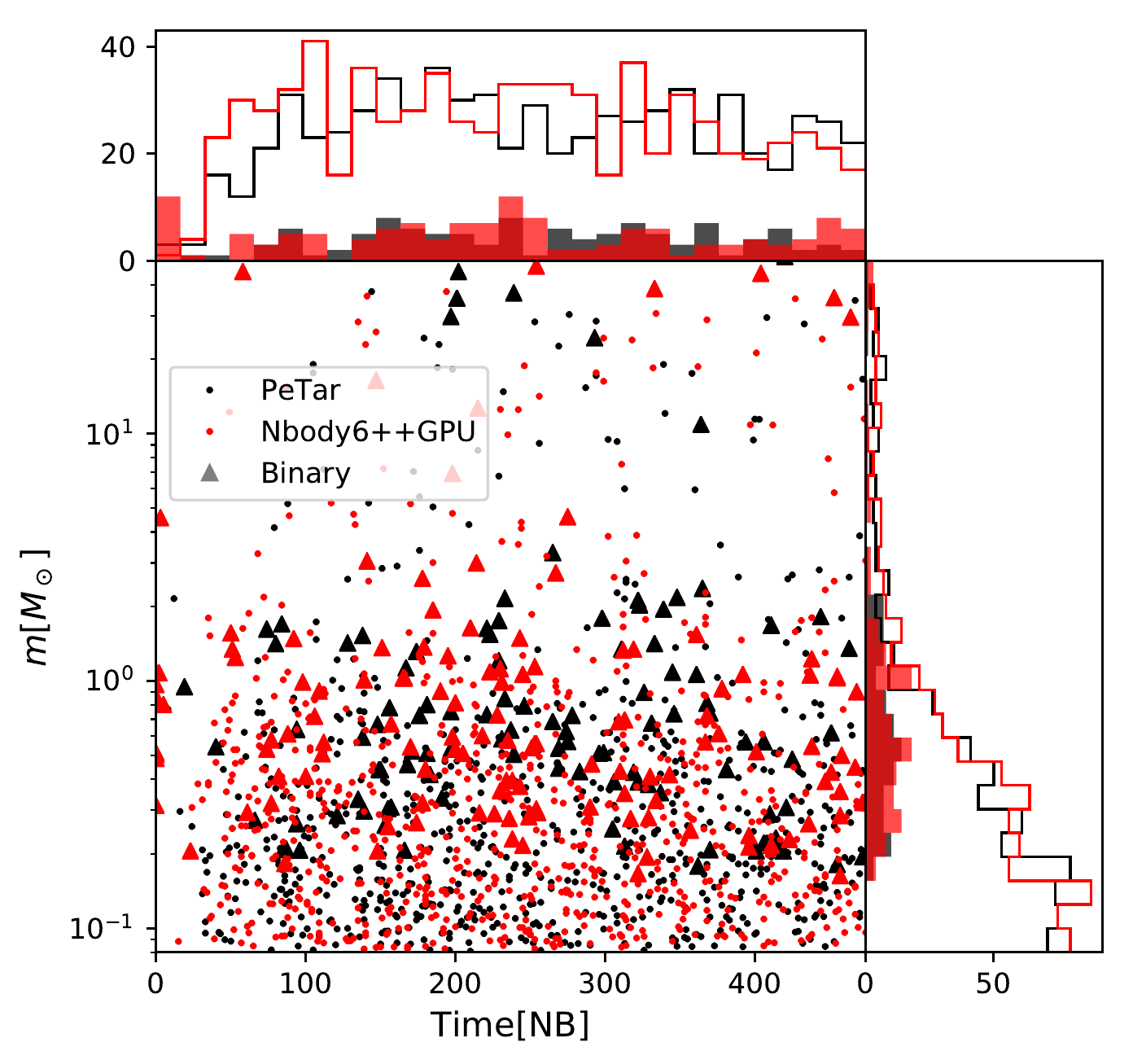}
  \caption{Distributions of the times and masses for single (dot) and binary (triangle) escapers in the N100kb model performed by using \petar~and \nbpp. 
    In the top and right histograms, single and binary escapers are represented by the step and bar styles, respectively.}
  \label{fig:N100kbesc}
\end{figure}

\begin{figure}
  \centering
  \includegraphics[width=1.0\columnwidth]{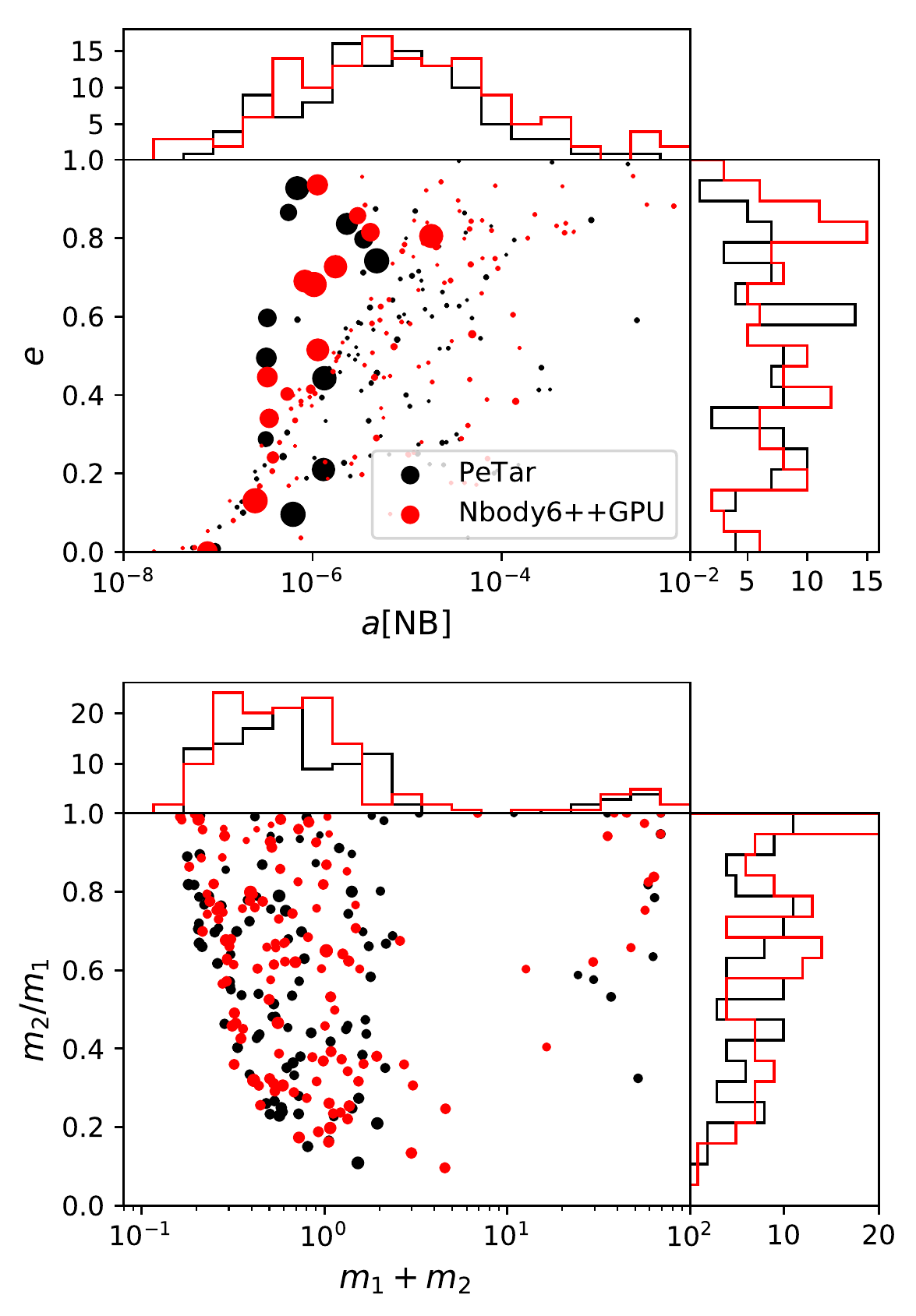}
  \caption{Upper: the distributions of semi-major-axis, $a$, and eccentricity, $e$ for binary escapers in N100kb models.
    The areas of markers are proportional to the binary masses.
    Lower: the distributions of the components' mass ratios and binaries' masses ($m_1+m_2$).
    The areas of markers are proportional to the logarithmic of $a$ with an offset.
  }
  \label{fig:N100kbescb}
\end{figure}

We also compare the properties of escapers in Fig.~\ref{fig:N100kbesc} and~\ref{fig:N100kbescb}.
Fig.~\ref{fig:N100kbesc} show the time and mass distribution of single and binary escapers.
The semi-major axis ($a$) vs. eccentricity ($e$) and component mass ratio ($m_2/m_1$) vs. binary mass ($m_1+m_2$) are shown in Fig.~\ref{fig:N100kbescb}.
The two codes well agree with each other on all properties compared here.
After $474$ NB time unit, the numbers of single and binary escapers are $689$ and $110$ in the case of \petar, and $733$ and $137$ in the case of \nbpp, respectively.
\nbpp~produce a slightly more escapers, but generally they agree well considering the statistical fluctuation. 
Therefore, \petar~and \nbpp~have a consistent treatment on the long-term dynamical evolution of binaries in star clusters.

\begin{figure}
  \centering
  \includegraphics[width=1.0\columnwidth]{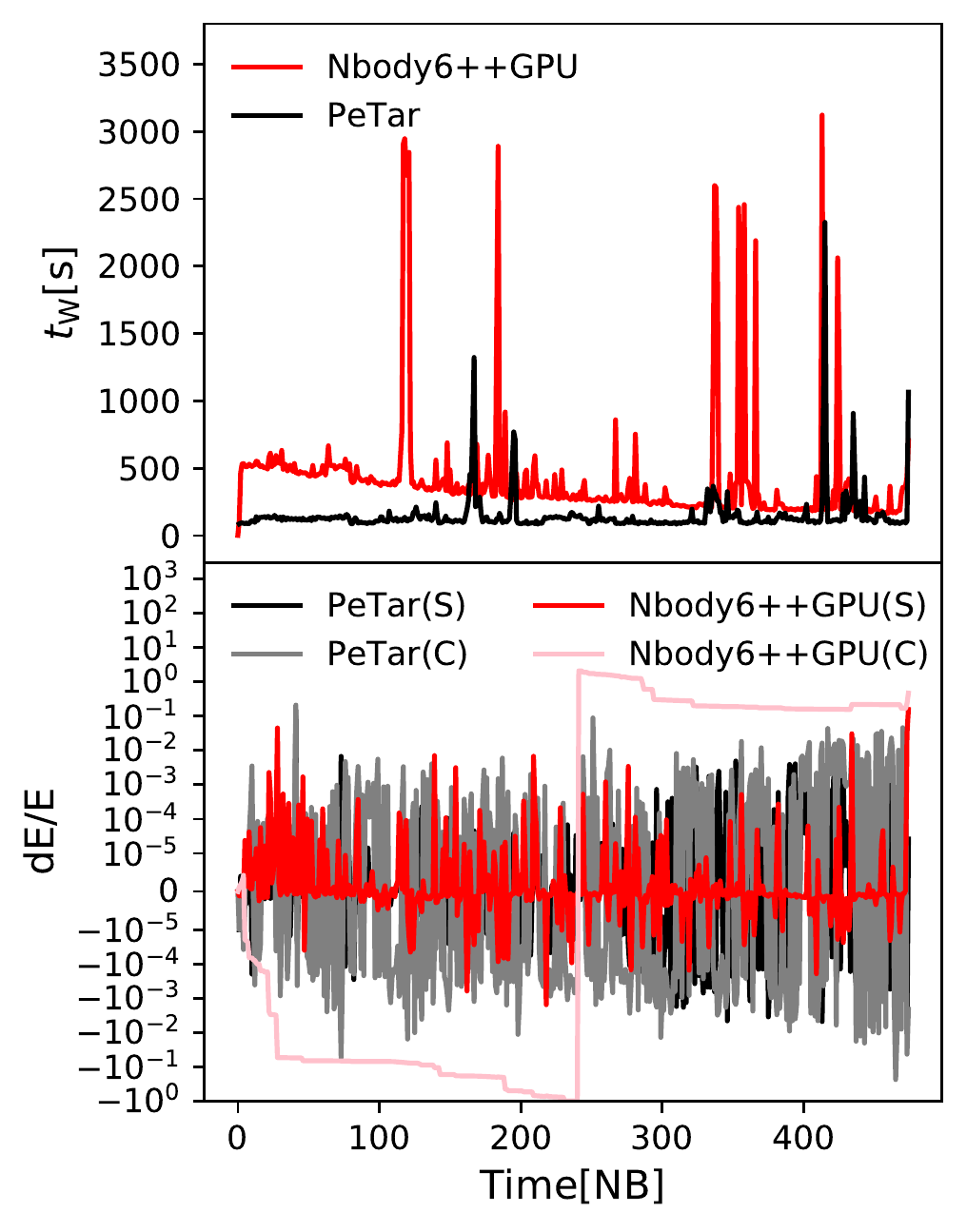}
  \caption{$\tcomp$ (upper panel) and $\diff E/E$ (lower panel) vs. the physical evolution time for the N100kb model. The plotting style is similar to Fig.~\ref{fig:perflong}.}
  \label{fig:perflongb}
\end{figure}

The performance and the energy error of the codes are shown in Fig.~\ref{fig:perflongb}.
Initially, \petar~is about 4 times faster than \nbpp~due to a more efficient treatment of primordial binaries.
The performance is stable till the end except for a few peaks due to the formation of stable multiple systems.
The performance of \nbpp~becomes better in the late time due to the disruption of wide binaries.
There are also several peaks where the performance significantly drops due to the appearances of stable multiple systems.
At the end, the performance of \petar~is about $1-2$ times faster than that of \nbpp.
Notice that here we compare a model of $10^5$ particles.
Due to the time consuming calculation in \nbpp, we have not compared a long-term evolution of million-body systems with large primordial binaries.
Based on the result of Fig.~\ref{fig:perfdesk}, we expect that the performance difference is more obvious in the case of simulating massive GCs.

\subsection{Scaling on supercomputer}

We test the performance of \petar~depending on $N$ and the number of cores ($\Ncore$) on the Cray XC50 supercomputer.
Each computing node has two of Intel Xeon Gold 6148 processors (Skylake), i.e. 40 cores.
The maximum of $\Ncore=960$.
No GPU devices are available.
The MPI and OpenMP are used together.
For $\Ncore > 10$, the number of OpenMP threads are fixed to be $10$ while the number of MPI processes is $\Ncore/10$.
For $\Ncore \le 10$, we only use OpenMP.
The AVX-512 acceleration is used.

The initial properties of the star clusters except $N$ are the same as that of the single and binary models used in Section~\ref{sec:perfdesk}.
The naming style of the models follows the style of N10k, N1m, N10kb, etc.
The configuration of input parameters follows Table~\ref{tab:inpar}.
Since $\Ncore$ is large, we set the maximum of $N$ to be 10 million.
The N10m and N10mb models have $\dts=1/2048$ and $1/4096$ and $\routr\approx0.00199303$ and $0.000996758$, respectively.
We choose the $\theta$ of $0.5$.

\begin{figure}
  \centering
  \includegraphics[width=1.0\columnwidth]{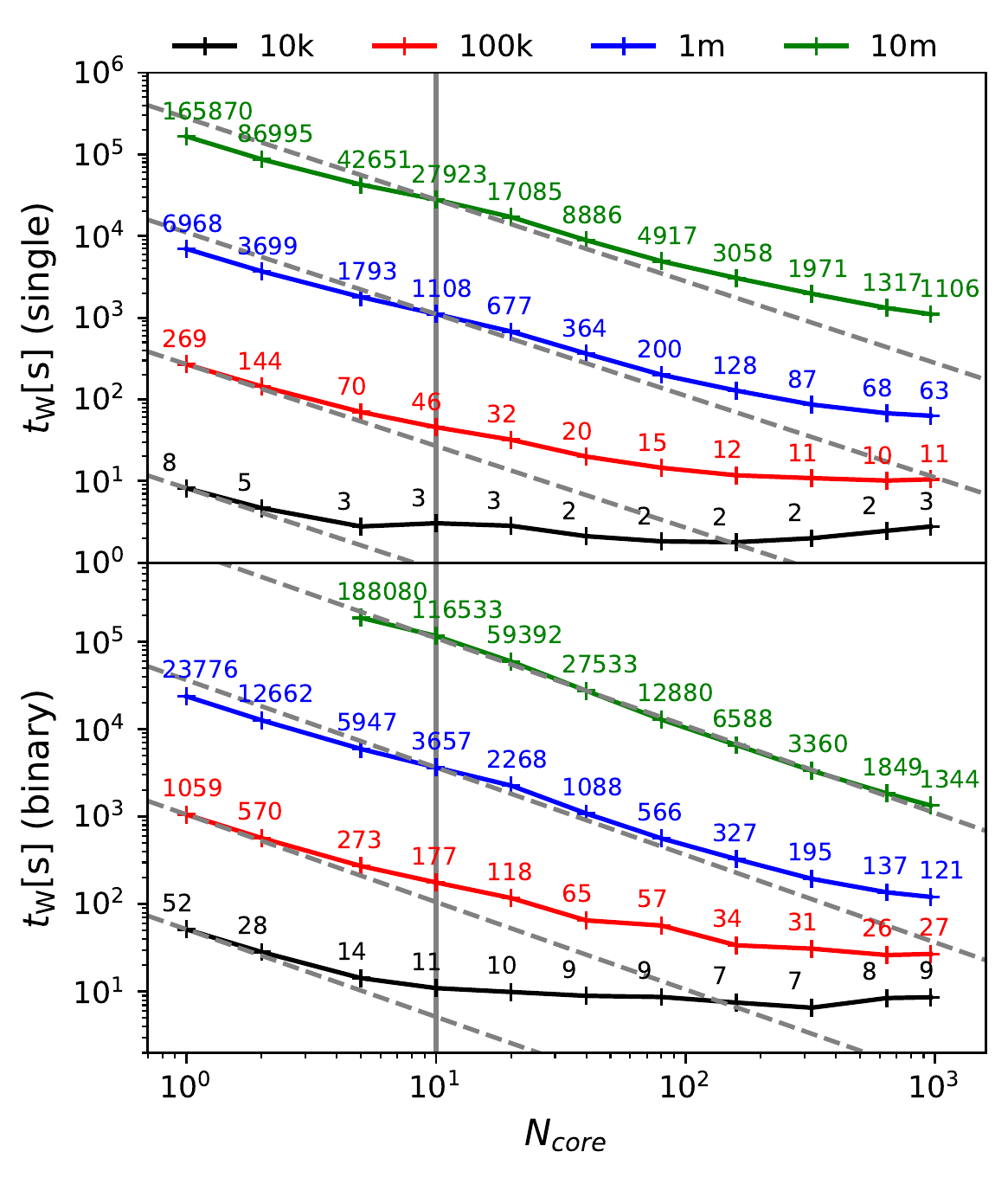}
  \caption{The performance test of \petar~on the Cray XC50 supercomputer.
    The two panels show $\tcomp$ of single and binary models vs the number of CPU cores ($\Ncore$), respectively.
    Colours indicate different $N$. The values $\tcomp$ in sec are shown above the data points.
    The dashed lines indicate the ideal scaling ($\propto 1/\Ncore$).
    The vertical line ($\Ncore=10$) indicates the boundary where the simulations use only OpenMP in the left and use both MPI and OpenMP in the right.
  }
  \label{fig:perfxc50}
\end{figure}

The result is shown in Fig~\ref{fig:perfxc50}.
The scaling of the N10k and N10kb model becomes flat when $\Ncore>5$.
The thresholds are about 160 and 640 for the case of $N=10^5$ and $10^6$, respectively.
The N10m and N10mb models can well scale up to $\Ncore=960$.
The absolute values of $\tcomp$ at the thresholds for million-body cases are about $63$ and $121$ sec for single and binary models, respectively.
The binary models roughly cost twice $\tcomp$ of that in single models, because for a given $N$, $\dts$ in the binary model is half of that in the single model.

The million-body DRAGON models with only $5\%$ primordial binaries took about $3000$ sec per NB time unit on a GPU-based supercomputer \citep{Wang2016}.
Thus, $\tcomp$ of the N10mb model are even much less than that of the DRAGON models.
Notice that we cannot directly compare the absolute $\tcomp$ as the Hardware are very different.
Besides, the long-term behaviour can be different from the initial case.
But we can still obviously see the significant advance of the performance by using \petar~on a modern supercomputer even without GPU.
Although in our test, the stellar evolution is not included, it should not significantly change the result as the computational complexity of that part is only $O(N)$.
It is expected that \petar~may even be possible to solve the ten-million-body problem.

\section{Conclusion and future work}
\label{sec:conclusion}

In this work, we have detailed described the new $N$-body code, \petar.
It is designed to simulate gravitationally bound collisional stellar systems with many subsystems.
The code is publicly available on GitHub (see footnote 1).
The hybrid integrator is implemented, where the long-range interaction is calculated by using the Barnes-Hut particle-tree method and short-range (neighbour) interactions are handled by the fourth-order Hermite with block time steps and the SDAR methods.
This hybrid method not only provides a calculation cost of $O(N \log{N})$ for the long-range interaction compared to $O(N^2)$ in the direct $N$-body method, but also can accurately follow the long-term evolution of perturbed binaries and hierarchical systems.

We introduce the mass-dependent changeover function for handling the systems with a wide range of particle masses (Section~\ref{sec:changeover}).
The clustering method based on the implementation in \textsc{pentacle} is improved by using mass and velocity dependent searching radii.
The orbit-dependent neighbour criterion is introduced to handle the high-velocity particles (Section~\ref{sec:clustering}).
In order to accurately follow the long-range perturbation to tight binaries when $\tbin<\dts$, the artificial particle algorithm is developed (Section~\ref{sec:tt}).

For a high performance on multiple-core computers, the code is implemented by using of \fdps, which provides a well optimized MPI and OpenMP parallelization of particle-tree (PT) part for the tree construction and long-range force calculations.
The neighbour searching and long-range force calculation in \pkick~are optimized by using of SIMD instructions (AVX, AVX2, AVX-512) and GPU acceleration (CUDA).
The \sdar library is used to perform the \pdrift~(short-range interactions).
This part is parallelized by using of OpenMP in each MPI processes.

A series of simulations are performed by using \petar~and \nbpp~in order to compare the performance and to validate whether \petar~can properly follow the long-term evolution of star clusters.
On a highly configured GPU based desktop, the performance of \petar~follows the scaling of $O(N\log{N})$ and is faster than \nbpp~(Fig.~\ref{fig:perfdesk}).
Especially for million-body systems with a large fraction of primordial binaries, \petar~can give an $11$ times faster performance.
Notice that the performance of \textsc{petar} is very sensitive to $\dts$ (Fig.~\ref{fig:finddt}). 
The best value is chosen in our test.

The test of \petar~on the Cray XC50 supercomputer show a good scaling depending on the number of CPU cores for $N\ge10^6$ (Fig.~\ref{fig:perfxc50}).
Million-body simulations with $100\%$ primordial binaries only take about $121$ sec wall clock time per NB time unit.
The 10 million models with and without ($100\%$) primordial binaries take $1344$ and $1106$ sec, which is even faster than that of the DRAGON models with only $5\%$ binaries.
Such significant improvement is due to the benefit of using new algorithms (\pppt~with SDAR), well optimized parallelization from \fdps and the advance of Hardware.

We also compare the long-term behaviour of the performance and relative energy error in Fig.~\ref{fig:perflong} and~\ref{fig:perflongb}.
In the case without primordial binaries, after core collapse, \petar~becomes slower due to the increasing of central density.
This is possibly improved if $\dts$ and $\routr$ are readjusted during or after the core collapse.

Sometimes, the computation becomes very slow during a short physical time interval of the models due to the formation of a specific type of stable hierarchical systems.
Such behaviours exist for both \petar~and \nbpp~in the models with primordial binaries (N100kb).
Once it happens, the benefit of parallel computing is lost due to a bad load balance.
However, how to efficiently and accurately solve such systems is a long existing question.
In the future work, the code will be improved to well handle a part of them.
But it is difficult to find a universal solution for all cases.
Because of this, the actual computing time of a long-term simulation may be longer than the prediction estimated from a few initial steps if such systems frequently form.

The results of N250k and N100kb models indicate that for both star clusters with and without primordial binaries, \petar~can provide a good agreement on the long-term behaviours of Lagrangian and core radii (including the post-core-collapse evolution; Fig.~\ref{fig:lagr} and~\ref{fig:lagrb}), the properties of single and binary escapers (Fig.~\ref{fig:N100kbesc} and~\ref{fig:N100kbescb}) and the evolution of binary orbits (Fig.~\ref{fig:binN100kb}).
Thus, \petar~is accurate enough to handle the realistic models of star clusters with many multiple systems.

The API to the stellar evolution package based on the framework of \textsc{sse} and \textsc{bse} \citep{Hurley2000,Hurley2002} are implemented.
The interface is designed in a way that switching between different versions of the \textsc{sse} and \textsc{bse} is straightforward.
Moreover, \textsc{petar} is also implemented as a module in a hydro dynamics code, \textsc{asura-bridge} (private comm. with Michiko Fujii), using the BRIDGE method \citep{Fujii2007}.
Thus, this hybrid code is possible to be used for studying the formation of star clusters, where few-body dynamics including close encounters, formation and evolution of binaries and hierarchical systems can be treated accurately.
Besides, the API to \textsc{amuse} \citep{PZ2013} is under developing.
The complete \petar~module in \textsc{amuse} will allow us to use BRIDGE to combine modules like hydrodynamic codes for a wide range of studies and to work with different single and binary stellar evolution packages.
On the other hand, the current version of \petar~only includes the pure Newtonian gravitational pair interaction.
In the future, the general relativity effect for compact object binaries including BHs and NSs will be implemented.

\section*{Data Availability}

The (benchmark) data underlying this article were generated by using the \textsc{petar} code on the desktop computer of the corresponding author and the supercomputer, Cray XC50 at Center for Computational Astrophysics (CfCA), National Astronomical Observatory of Japan. The data underlying this article will be shared on reasonable request to the corresponding author.
The code, \textsc{petar}, introduced in this article is publicly available in GitHub at \petarurl, under the MIT license. 

\section*{Acknowledgements}

We thank Sverre Aarseth for the helpful suggestions on the code development.
L.W. thanks the financial support from JSPS International Research Fellow (School of Science, The university of Tokyo) and the support from Alexander von Humboldt Foundation (The University of Bonn).
Numerical computations were in part carried out on Cray XC50 at Center for Computational Astrophysics, National Astronomical Observatory of Japan.

\bsp

\label{lastpage}

\end{document}